\documentclass[crop]{CSL}%%%%where CSL is the template name

\usepackage{natbib}
\usepackage{hyperref}
\usepackage{graphicx}
\usepackage{bbm}
\usepackage{physics}
\usepackage{subcaption} % for subfigures
\usepackage{tabularx}
\usepackage{booktabs}
\usepackage{xcolor}
\usepackage{comment}
\usepackage{csquotes}
\usepackage{txfonts}

\newcolumntype{Y}{>{\centering\arraybackslash}X}  % Center-Aligned column marker for tabularx2

\newcommand{\citewy}[1]{%
  \citeauthor{#1}%
}

\begin{document}

\supertitle{Research Paper}

\title[A revision of the lifetime of submoons]{A revision of the lifetime of submoons: tidal dynamics with the Euler-Lagrange equation}

\author[Saganas \textit{et al.}]{Iason Saganas$^{1}$, Grant Mayberry$^{2}$, Barbara Ercolano$^{3,4}$}

\address{\add{1}{University Observatory Munich, Faculty of Physics, Ludwig-Maximilians-University Munich, Scheinerstr. 1, 81679 Munich, Germany}, \add{2}{Vanderbilt University, 2201 West End Ave, Nashville, TN 37235, USA}, \add{3}Exzellenzcluster “Origins”, Boltzmannstr. 2, D-85748 Garching, Germany and \add{4} Max-Planck-Institut für Extraterrestrische Physik, Giessenbachstra.e 1, 85748 Garching, Germany}

\corres{\name{Iason Saganas} \email{I.Saganas@campus.lmu.de}}

\begin{abstract}

Submoons, moons orbiting other moons, may be exotic environments capable of hosting extraterrestrial life. We extend previous studies to revise the maximum lifetime of these objects due to planetary, lunar and sublunar tidal migration. 

Using the Euler-Lagrange equation with a tidal dissipation process as specified by the Constant Geometric Lag model, we derive and solve the governing equations numerically to map the semi-major axis parameter space for star-planet-moon-submoon systems in which the submoon could be massive enough to host life.

% To find the differential equations governing the four-nested-body system, we make use of the Euler-Lagrange equation with a non-zero right hand side, using the tidal energy specified by the Constant Geometric Lag model in order to account for the dissipative tidal process. We solve the resulting equations numerically to explore the semi-major axis parameter space, searching for star-planet-moon-submoon configurations, in which the submoon is massive enough to potentially host life itself.

We find that Earth could have hosted asteroid-sized submoons ($\sim10^{15}\mathrm{kg}$), whereas a submoon near the previously proposed upper limit ($\sim4.6\cdot10^{17}\mathrm{kg}$) would have driven the Moon $\sim30\%$ farther from Earth than its current orbit.

% Our results show that Earth could have hosted an asteroid-sized submoon of $\sim 10^{15}\mathrm{kg}$ in its current configuration. A previous analysis found an upper-bound of $m_{sm}\sim 4.6\cdot 10^{17}\mathrm{kg}$ for Earth's current configuration, while our numerical analysis suggests that, for such a submoon mass, the Moon would have migrated approximately $30\%$ of its current distance farther outward.  

A Warm Jupiter system like Kepler1625 has greater potential of hosting a massive submoon. We found that a submoon of around $10\%M_{\text{Luna}}$ could survive if Kepler1625b's hypothesized moon were $68\%$ farther away then what the best-fit model suggests ($67R_{\mathrm{p}}$ instead of $40R_{\mathrm{p}}$). Giant submoons of mass $1.8M_{\oplus}$ are stable in a Kepler1625-like system. In these cases, the moon orbit is wide ($> 100R_{\mathrm{p}}$). Decreasing the submoon mass to a habitability prerequisite of $0.5M_{\oplus}$, likely needed for a stable atmosphere and plate tectonics, leads to a smaller total number of stable iterations relative to the $m_{sm}=1.8M_{\oplus}$ case.

In fact, we identified a minimum number of stable iterations on intermediate submoon mass-scales of around $0.1M_{\oplus}$. This is likely due to an interplay between small tidal forces at small submoon masses and small Roche-Limits at very high submoon masses. If submoon formation pathways in Warm Jupiter systems prefer such intermediate mass-scales, habitable submoons could be a rare phenomenon. 

\end{abstract}

%\keywords{Euler–Lagrange equation, habitability, orbital stability, submoons, tidal energy, tidal migration, warm jupiters}  % keywords will be added during the submission process

  % needed empty for Fpagebreak to work

\selfcitation{Saganas I \textit{et~al.} (2025) A revision of the lifetime of submoons: tidal dynamics with the Euler-Lagrange equation. https://doi.org/xxxxx}

\received{xx xxxx xxxx}

\revised{xx xxxx xxxx}

\accepted{xx xxxx xxxx}

\maketitle
\Fpagebreak

\section{Introduction}

The question of life beyond Earth naturally invites speculation about the diversity of worlds that potentially harbour extraterrestrial life. Not only exoplanets, but large exomoons around giant gas planets could host life; this possibility was explored as early as 1997 by \citewy{Williams1997}, and was recently extended to exomoons orbiting free-floating planets (e.g. \cite{Avila2021}; \cite{Roccetti2023}) % DON'T SET "."

In a paper that renewed interest in the topic, \cite{Kollmeier_2018} examined the maximal lifetime of moons of moons ("submoons") as potential hosts for life. This was done by exploring the minimum radius a moon must have, given its density, to host a submoon of a fixed mass over the time span of $4.6\mathrm{Gyr}$.

Through an analytical estimation considering only moon-submoon-tides, based on the work of \cite{Barnes_2002}, they found that Jupiter, Saturn, Earth, as well as Kepler-1625b could have hosted a small submoon on the order of $10^{-6}M_{\text{Luna}}$, corresponding to a spherical object of $20\mathrm{km}$ radius (assuming a submoon bulk density of $2\mathrm{g/cm^3}$). Due to its hypothetical giant moon, Kepler-1625b is a suitable environment for a more massive submoon. \cite{Kollmeier_2018} found that, in its current configuration in parameter space, Kepler-1625b would be able to host a $\sim 10^{-3}M_{\text{Luna}}$ submoon\footnote{See \href{https://github.com/iason-saganas/stability-of-submoons/blob/master/media/original-submoon-paper/translation_of_IDL_file.py}{\texttt{python} script}} at most.

   In this paper, we want to extend the work by \cite{Kollmeier_2018} by considering not only the tides acting between a moon and its submoon, but also the ones acting in the planet-moon- and planet-star-subsystem simultaneously. 

   This allows us to consider effects on the submoon's lifetime caused by the tidal dynamics in e.g. the planet-moon-subsystem: If the moon crashes into the planet, the submoon's lifetime is cut short, if the moon escapes from the planet, its Hill-Radius increases, delaying a potential escape of the submoon from the system and thereby increasing its longevity.  

Our approach consists of finding the Euler-Lagrange equation for the tidal dynamics described via the so-called "Constant Geometric Lag" (CGL) model for a general $n$-nested-body-system, assuming coplanarity and circularity of all orbits. Upon evaluation for $n=4$, we can extract the differential equations that govern the system. We solve those equations numerically using \texttt{SciPy}'s \citep{SciPy2020} \texttt{solve\textunderscore ivp} routine. In particular, we formulate an initial value problem by iterating over all possibly stable configurations of the star-planet-moon-submoon system, for which the Roche-Limit (disintegration of a body) and the critical semi-major axis (escape from the system) need to be taken into consideration as lower and upper bounds, respectively.

While the analytical approach of \cite{Kollmeier_2018} allowed them to estimate the maximum submoon mass in existing systems like the Sun-Earth-Moon system, our formulation enables us to determine the optimal configuration of an Earth-like system that would allow a Moon-like satellite to host the most massive submoon permitted by our assumptions. For this, we assume a mass ratio of $1/10\text{th}$ between an orbiting body and its host.

Deriving the differential equations of interest from the Euler-Lagrange equation is not only of intellectual interest but also facilitates future adaptations to more complex orbit geometries, since the formulation is based on scalar functions rather than vector-valued quantities like torque.

The tidal potential needed for the derivation is based on the widely-used Constant Geometric Lag model as specified, e.g., by \cite{Murray_Dermott_2000}, chap. 4.9. In this model, the tidal response is modeled through a forced harmonic oscillator that leads to a constant phase shift in the tidal response of the planet. The model under consideration exhibits mathematical limitations (see second-to-last section), so our quantitative results are primarily indicative and await refinement through the incorporation of more realistic tidal models. 

In the past, the orbital constraints of possible submoons in the Kepler1625 system have been studied based on $N$-body simulations  \citep{RosarioFranco2020SubmoonNBodies}. 

For a treatment of a star-planet-moon system using the \enquote{Constant Time Lag Model}, see \cite{Piro_2018}.

This work is structured as follows. After this introduction, we summarize the basic theory of tidal evolution of satellites based on the Constant Geometric Lag model. In the following section, we derive the energies needed for the Euler-Lagrange equation subject to our assumptions (circularity and co-planarity of all orbits) and the specific tidal model and derive the differential equations of interest. Aspects of numerical stability and the results of the simulations are presented, before touching upon limitations and summarizing this work.

\section{Basic theory}\label{sec: Basic Theory}

\subsection{Tidal migration}

 Tidal effects occur when taking into account that the gravitational potential of celestial bodies has a point-particle contribution only up to first order. In reality, a planet has some volume and can be thought of being made up of a collection of infinitesimal mass packets $\mathrm{d}m$, resulting into one side of the planet experiencing gravity more strongly than the other. The result of this interaction is a gravitational gradient that builds up accross the planet's surface and rearranges the positions of the little mass packets $\mathrm{d}m$, such that a tidal bulge forms (see Fig. \ref{fig: Coordinate system definition}).   
   
   The formation of the tidal bulge breaks the azimuthal symmetry of the system, eventually leading to an exerted gravitational torque that transfers energy and angular momentum, affecting the planet's and moon's spin rate $\Omega_{\mathrm{p}}$, $\Omega_{\mathrm{m}}$ and the moon's orbit frequency $n_{\mathrm{m}}$. 
   
   By Kepler's Law, 
\begin{equation}
    n = \mu^{\frac{1}{2}}a^{-\frac{3}{2}},
\end{equation}
a change in the orbital angular momentum must be balanced by a change in semi-major axis, $a$, leading to "tidal migration". Here, $\mu=G(m_{i}+m_j)$ is the "standard gravitational parameter" of two bodies indexed by $i$ and $j$. 

Numerous tidal pathways have been discovered numerically and analytically, e.g., in the two-body case \citep{Counselman_1973}:

% \cite{Sasaki_2012} analyzed a star-planet-moon system in which the planet is torqued by stellar and lunar tides. 

% They refer to the work of \cite{Counselman_1973}, who found two dynamic and one static evolution state in the base system of a satellite orbiting a planet:

\begin{itemize}
    \item The satellite's semi-major axis decreases up to the point where it is disintegrated at the Roche-Limit $a_{\text{Roche}}$
    \item The satellite's semi-major axis increases until the satellite escapes the gravitational influence of the host body ($a_{\text{crit}}$)  
    \item The host's spin angular and the satellite's orbital angular momentum enter a stable resonance.
\end{itemize}

While including an additional body like a star increases the variety of possible evolution states\footnote{In this study, we do not categorize our numerical solutions as, e.g., \cite{Sasaki_2012} do in their work. The latter thus serves as a foundation for understanding the physical scenarios suggested by numerical simulations explicitly.} \citep{Sasaki_2012}, two important boundary conditions become clear from the two-body system. First, the Roche-Limit, 
\begin{equation}\label{eq: Roche limit}
    a_{\text{Roche}, \mathrm{m}} =  R_{\mathrm{m}} \cdot\bigg(\frac{3m_{\mathrm{p}}}{m_{\mathrm{m}}} \bigg)^{\frac{1}{3}}
\end{equation}
\citep{Murray_Dermott_2000}, as a lower bound for a satellite's semi-major axis, with $m_p$ and $m_m$ the planet's and moon's mass respectively. Secondly, the so-called critical semi-major axis $a_{\mathrm{crit}}$, which gives the upper bound after which an orbiting body escapes the gravitational influence of its hosting body. Numerical simulations have found $a_{\text{crit}}$ to be a constant fraction of the hosting body's Hill-Radius: 
\begin{equation}
    a_{\text{crit}} = \begin{cases}
        0.4 a_{\mathrm{Hill}} \text{\hspace{7mm} for a moon}\\ 
        0.33 a_{\mathrm{Hill}} \text{\hspace{5mm} for a submoon} 
    \end{cases}
\end{equation}
\citep{RosarioFranco2020SubmoonNBodies}. In a star-planet-moon-system, the Hill-Radius relevant to the moon is
\begin{equation}
    a_{\text{Hill}, \mathrm{m}} =  a_{\mathrm{p}}\cdot  \bigg( \frac{m_{\mathrm{p}} }{ 3m_{\mathrm{s}}}\bigg)^{\frac{1}{3}}
\end{equation}
\citep{Sasaki_2012}. The dependence is again on the planet's mass $m_p$, as well as the star's mass $m_s$, giving a measure of the region where the gravitational influence of the planet dominates, rather than the star's.  

\subsection{Non-central potential and geometric lag}\label{subsec: Non-Central potential and Geometric Lag}
  
\begin{figure*}
   \centering
   \includegraphics[width=0.9\linewidth]{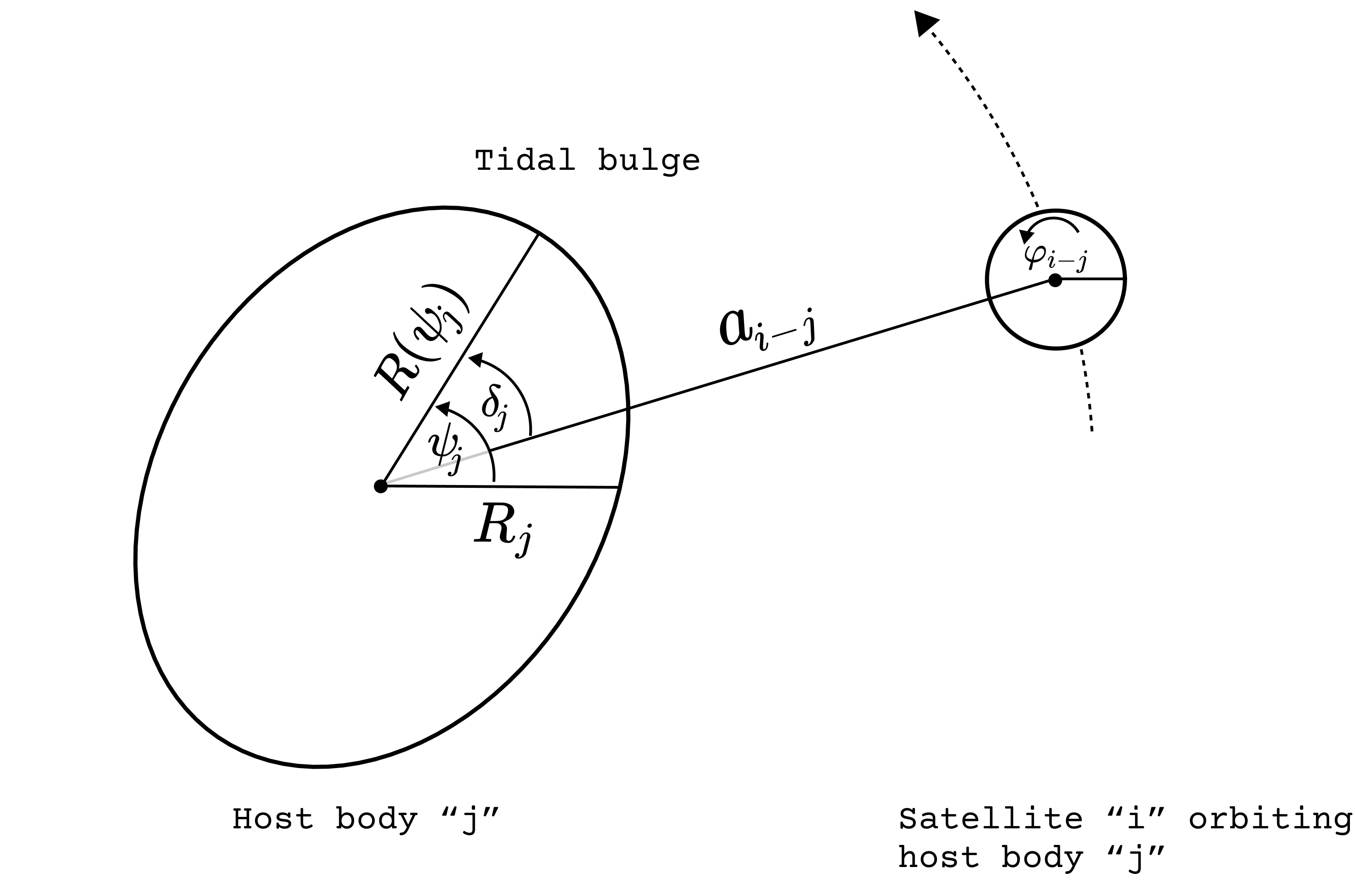}
   \caption{A satellite tidally disturbing its host. The tidal bulge forms at a location that is shifted by $\delta_j$ with respect to the position of the tide-raising satellite on the sky. In this case, the host's spin rate exceeds the satellite's orbital frequency, leading to the mass concentration being carried ahead of the satellite. It is the mass concentration at this angle that has the most important contribution to the non-central exterior potential and thus it enters in the tidal energy, Eq. \eqref{eq: i-j-th Tidal energy}. The little mass packets making up the tidally disturbed body are characterized by the surface angle $\psi_j$ with the change in this angle being the host's spin frequency $\Omega_j$.}
              \label{fig: Coordinate system definition}%
\end{figure*}

A small mass element on the surface of a deformed planet generates a gravitational potential in which a point-like satellite moves (see Fig. \ref{fig: Coordinate system definition}). The mass packet's location can be characterized by a surface angle $\psi$, that is measured from an arbitrary reference line, such that $\dot{\psi}=\Omega$ is the spin frequency of the planet. The angle measured between the mass packet at $\psi$ and the position of the satellite in the sky ($\varphi$) is, in general, a time-dependent quantity, since the spin frequency of the planet $\Omega$ and the orbit frequency of the satellite $\dot{\varphi}=n$ do not have to be equal. This generic, time-dependent phase shift, denoted by $\delta$, can be calculated as: 
 \begin{equation}\label{eq: Generic delta shift}
     \delta = \psi - \varphi - \pi.
 \end{equation}
 %(see Fig. \ref{fig: Coordinate system definition}). 
 % From here on out we drop the phase shift of $\pi$, since this will cancel out in the later derivatives. 
In the Constant Geometric Lag model, the deformation of the planet due to the potential generated by its satellite is modelled as a driven harmonic oscillator. Dissipation effects lead to a constant offset between where the tidal bulge forms (at an angle $\psi_T$) and the position of the tide-raising satellite on the sky $\varphi$. The position of the tidal bulge, $\psi_T$, is a special point on the planet's surface: Since the offset between the tidal bulge and tide-raising satellite is set to be a constant, $\delta_{\mathrm{lag}}$, with
 \begin{equation}\label{eq: Definition of lag angle via psi_T}
     \delta_{\mathrm{lag}} = \psi_T-\varphi- \pi, 
 \end{equation}
the velocity associated with the point $\psi_T$ must be 
\begin{equation}
    \dot{\psi}_{\mathrm{T}} = \dot{\varphi}=n,
\end{equation}
since only then is $\dot{\delta}_{\text{lag}}=\dot{\psi}_{\mathrm{T}}-\dot{\varphi}=n-n=0$. The non-uniform mass concentration around $\psi_T$ generates a non-central potential in which the satellite moves, effectively creating a celestial gravitational wrench that over billions of years acts to reduce the lifetime of the satellite. 

This lag angle $\delta_{\text{lag}}$ is positive if the planet spins faster than the satellite orbits it, $\Omega > n$, and negative otherwise. Throughout \cite{Murray_Dermott_2000}, chapter 4,  it is implied that the absolute value of the tidal lag angle $\delta_{\text{lag}}$ times two is equal to the reciprocal of a "quality factor" $Q$ - A difficult to access quantity that parameterizes the interior dissipation process. Often, only vague estimations of this quantity can be made. In summary:
\begin{equation}\label{eq: Lag angle and quality factor}
    \mathrm{sin}(2\delta_{\text{lag}}) \approx 2\delta_{\text{lag}}= 2\mathrm{sgn}(\Omega-n)Q^{-1},
\end{equation}
where $\mathrm{sgn}$ is our notation for the signum function of the relative tidal forcing frequency $\Omega-n$.

The gravitational potential generated by a generic point on the planet's surface can be described as  
\begin{equation}\label{eq: Basic non-central potential before expansion}
     V_{\text{tidal}} = \frac{G\int \mathrm{d}m_p}{\Delta(\delta)},
 \end{equation}
where $\Delta (\delta)$ is the distance between the location of a small mass element $\mathrm{d}m_p$ and the location of a moon moving in that potential. A Taylor-Expansion can show that the inverse distance $1/\Delta (\delta)$ can be approximately expressed as a series of "Legendre Polynomials" $\mathcal{P}_n$ in the generic phase shift angle $\delta$, characterizing the location of a specific mass packet on the planet's surface:
\begin{equation}
    \frac{1}{\Delta(\delta)} \approx \frac{1}{a_m}\sum_{n=0}^{\infty}\bigg(\frac{R_p}{a_m}\bigg)^n \mathcal{P}_n(\cos\delta)
\end{equation}
\citep[see][chap. 4.3]{Murray_Dermott_2000}. Here $R_{\mathrm{p}}$ refers to the planet's mean circular radius and $a_{\mathrm{m}}$ to the moon's semi-major axis (assuming $R_p \ll a_m$). 

By inputting  the expanded form of $1/\Delta(\delta)$ into Eq. \eqref{eq: Basic non-central potential before expansion} and carrying out the integration $\int \mathrm{d}m_p$ explicitly, taking into consideration a specific surface equation model ("radial tide" model, see chapter 4.3, Eq. 4.44 of \cite{Murray_Dermott_2000}), one finds 
\begin{equation}
   V_{\text{tidal}} = -G \frac{m_p^2}{m_m}k_{2m}\frac{R_m^5}{a_{m}^6}\mathcal{P}_2(\cos\delta)
\end{equation}
for the non-central gravitational potential that is responsible for the tides. This may be generalized to the case where a body $j$ is tidally disturbed by another body $i$ via the corresponding tidal energy as: 
\begin{equation}\label{eq: i-j-th Tidal energy}
    U_{\text{tidal}, i-j}=-Gm_i^2k_{2j}\frac{R_j^5}{a_{i-j}^6}\mathcal{P}_2(\cos\delta).
\end{equation}

\subsection{Functional relationship between variables}\label{subsec: Functional relationship between variables}

Note that in Eq. \eqref{eq: i-j-th Tidal energy}, the $\delta$ is - a priori - the generic phase shift given by Eq. \eqref{eq: Generic delta shift}, instead of the lag angle $\delta_{\mathrm{lag}}$ given by Eq. \eqref{eq: Definition of lag angle via psi_T}. The formalism is only to be evaluated at $\delta = \delta_{\mathrm{lag}}$ (or equivalently, $\psi=\psi_T$) after the Euler-Lagrange equation has been used to extract the relevant differential equations. Eq. \eqref{eq: Lag angle and quality factor} for the lag angle results only after having done an analysis of the harmonic oscillator model. Assuming it beforehand disrupts the functional relationship of the variables in the Euler-Lagrange equation. 

%Another example for this is the planarity of the two-body problem: In a barycentric frame of reference, the $z$-coordinate of the position vectors of both bodies can be fixed to $0$. Assuming this $z=0$ sub-solution to the Keplerian Problem in order to build the Lagrangian of the system leads to the expected, correct differential equations, but is in general bad practice \citep[see note on page 133]{Fliessbach_2020}. Planarity is in this sense \enquote{well behaved}. 

Another example of such a "disruption" is the usage of Kepler's Law
% An example of \enquote{bad behaviour} in this sense is the usage of Kepler's Law
\begin{equation*}
    n = \mu^{\frac{1}{2}}a^{-\frac{3}{2}}.
\end{equation*}

Assuming this relationship a priori and then taking the necessary time, position and velocity derivatives in the Euler-Lagrange equation will lead to false results\footnote{Injecting known sub-solutions into a problem a priori is bad practice, see also note on page 133 of \cite{Fliessbach_2020}.}.

So, to properly capture the relationship between all variables, $\delta=\psi - \varphi-\pi$ needs to be considered for the tidal energy, representing the generic phase shift of a potential point $\psi$ at which the tidal bulge could form. Only after taking the necessary derivatives we evaluate at the concrete point $\psi = \psi_{\mathrm{T}}$ at which the tidal bulge actually does form, parameterized through the quality factor $Q$: 
\begin{equation}
    \frac{\partial \mathcal{P}_2(\cos\delta)}{\partial \psi}\bigg|_{\psi=\psi_T} \approx -\frac{3}{2}\cdot (2\delta)\bigg|_{\psi=\psi_T} \stackrel{\eqref{eq: Lag angle and quality factor}}{=} -\frac{3}{2}\mathrm{sgn}(\Omega - n)Q^{-1},
\end{equation}
\begin{equation}
    \frac{\partial \mathcal{P}_2(\cos\delta)}{\partial \varphi}\bigg|_{\psi=\psi_T} \approx +\frac{3}{2}\cdot (2\delta)\bigg|_{\psi=\psi_T} \stackrel{\eqref{eq: Lag angle and quality factor}}{=} +\frac{3}{2}\mathrm{sgn}(\Omega - n)Q^{-1},
\end{equation}

\section{Derivation of the Euler-Lagrange equation}\label{sec: Derivation of the Euler-Lagrange equation}

\subsection{Assumptions and notation}\label{subsec: Assumptions and notation}

In the following, we will assume circular, planar and prograde orbits of all bodies for simplicity. We will also neglect tidal effects of  \enquote{indirect} orbits (e.g. effect of stellar torque on moon). We will assume that the distance between a host and the barycenter of a system is negligible in comparison to the satellite's semi-major axis, which translates to a constraint of the mass ratio of satellite to host (e.g., $1/10\text{th}$, see following subsection).

In the equations that follow, we will index the variables and constants from Table \ref{table: Definition of variables and constants.} with subscripts that represent the star, planet, moon and submoon by \enquote{$\mathrm{s}$}, \enquote{$\mathrm{p}$}, \enquote{$\mathrm{m}$} and \enquote{$\mathrm{sm}$} respectively. 

We derive the Lagrangian of an arbitrarily nested planetary system consisting of $n$ bodies and evaluate the Euler-Lagrange equation afterward for $n=4$ to get a representation of a submoon system. To do this, we label the celestial bodies by their "host level" $h$, for convenience, which is simply the counting index when enumerating the bodies sequentially starting from the star (see Table \ref{table: Definition of the hierarchy number.}). The set of of all those numbers is $\mathcal{H}$:
\begin{equation}
    \mathcal{H}:= \{ \hspace{1mm} h \hspace{1mm} \mid \hspace{1mm} h \leq N \hspace{1mm} \},
\end{equation}
where $N$ is the number of bodies in the nested system.

Equations are formulated in terms of  "tidally disturbed bodies" (represented by the index $j$) and the associated "tidal disturber" (represented by the index $i$) and sums over the set of all host levels $i,j\in \mathcal{H}$ are constructed to get the Lagrangian of the system. 

Inserting the latter into the Euler-Lagrange equation, two free indices, $k$ and $l$, will remain, which label the generalized coordinates $q_{k-l}$. 

%. Conceptually, we may again think of these indices as representing a "tidal disturber" and the "tidally disturbed body", respectively (see Table \ref{table: Description of used indices ijkl.}). 

Finally, we can map the two free indices $k$ and $l$ to literal subscripts (like \enquote{m} and \enquote{p} for moon and planet, see Table \ref{table: Definition of the hierarchy number.}) to obtain concrete results.

\begin{table}[!h]
\footnotesize
\tabcolsep=4pt
\processtable{Definition of variables and constants. \label{table: Definition of variables and constants.}}{
\begin{tabularx}{\columnwidth}{YY}
\rowcolor{Theadcolor}
\textbf{Definition} & \textbf{Symbol} \\
Mass & $m$ \\
Standard gravitational parameter & $\mu = G(m_1+m_2)$ \\
Mean radius & $R$ \\
Semi-major axis & $a$ \\
Surface angle & $\psi$ \\
Spin frequency & $\dot{\psi} = \Omega$ \\
Orbit angle & $\varphi$ \\
Orbit frequency & $\dot{\varphi} = n$ \\ 
Second tidal Love number (measure of rigidity) & $k_2$ \\
Quality factor (dissipation parameter) & $Q$ \\
Moment of inertia & $I$ \\
Kinetic energy & $T$ \\
Gravitational energy & $U_{\text{gravi}}$ \\
Tidal energy & $U_{\text{tidal}}$ \\
Lagrangian & $\mathcal{L}$ \\
\end{tabularx}}{}
\end{table}

\begin{table}[!h]
\footnotesize
\tabcolsep=4pt
\processtable{Host levels in the four-nested-body system. \label{table: Definition of the hierarchy number.}}{
\begin{tabularx}{\columnwidth}{YYY}
\rowcolor{Theadcolor}
\textbf{Body} & \textbf{Host level} $h$ & \textbf{Subscript} \\
Star & 1 & $\mathrm{s}$ \\
Planet & 2 & $\mathrm{p}$ \\
Moon & 3 & $\mathrm{m}$ \\
Submoon & 4 & $\mathrm{sm}$ \\
\end{tabularx}}{}
\end{table}

\begin{table}[!h]
\tabcolsep=4pt
\processtable{List of used indices in the derivation of the Euler-Lagrange equation. 
%The index type and its is indicated. and purpose used to construct the total Lagrangian of the system in the Euler-Lagrange equation and free indices that determine which differential equation is extracted. All indices run over the set of all hierarchy numbers $\mathcal{H}$.
\label{table: Description of used indices ijkl.}}{
\begin{tabularx}{\columnwidth}{YYY}
\rowcolor{Theadcolor}
\textbf{Index} & \textbf{Represents} & \textbf{Type} \\
$i$ & Tidal disturber & Summation \\
$j$ & Tidally disturbed body & Summation \\
$k$ & Tidal disturber & Free \\
$l$ & Tidally disturbed body & Free \\
\end{tabularx}}{}
\end{table}

\subsection{Potential, kinetic and tidal energy}\label{subsection: Potential, Kinetic And Tidal Energy}

Let $j$ refer to a tidally disturbed body and $i$ to its tidal disturber. The total potential energy in this two-body system is the sum of the standard gravitational energy between the bodies and the gravitational energy associated with the non-central potential due to the matter redistribution caused by tides: 
\begin{align}\label{eq: Total Potential Energy}
    U_{\text{total},i-j} &= U_{\text{gravi},i-j}+U_{\text{tidal}, i-j} \nonumber \\
    &=-G\frac{m_im_j}{a_{i-j}}-Gm_i^2k_{2j}\frac{R_j^5}{a_{i-j}^6}\mathcal{P}_2(\cos\delta_j)
\end{align}
The total kinetic energy is composed of contributions by the rotational (rot.) energy,
%Modeling the tidal disturber as a point-particle with no inherent rotation around its own axis:
\begin{equation}\label{eq: Rotational kinetic energy}
    T^{\mathrm{(rot.)}}_{j} = \frac{1}{2}I_j \Omega_j^2,
\end{equation}
and orbital (orbit.) energy:
\begin{equation}\label{eq: Orbital Kinetic Energy}
    T_{i-j}^{\mathrm{{(orbit.)}}}=\frac{1}{2}\frac{m_im_j}{m_i+m_j}\big(\dot{a}^2_{i-j}+a^2_{i-j}n^2_{i-j}\big),
\end{equation}
such that the total kinetic energy in the subsystem is: 
\begin{equation}\label{eq: Kinetic Energy}
    T_{i-j} = T_{i-j}^{\mathrm{{(orbit.)}}} +T^{\mathrm{(rot.)}}_{i}+T^{\mathrm{(rot.)}}_{j}.
\end{equation}
The reduced mass of the two bodies appearing in Eq. \ref{eq: Orbital Kinetic Energy} emerges when assuming circularity and summing the kinetic energies in a barycentric frame of reference. Importantly, the orbital energy in the expression assumes that the distance between the $j$-th and $i$-th body is simply the semi-major axis of the orbiting body. This means that the distance between the larger body and the barycenter is considerably smaller than the mutual distance of the two bodies. 

Let $a_1$ and $a_2$ be the semi-major axes of two bodies $M$ and $m$ that orbit around their mutual barycenter. Demanding that the distance $r$ between the bodies, i.e. $r=a_1+a_2$, is approximately $a_2$ essentially amounts to requiring that $m/M\ll 1$. This can be seen by considering the formulas $a_1=r\cdot m/(m+M)$ and $a_2=r\cdot M/(m+M)$ in the limit $a_1/a_2 \ll 1$. For our purposes, we will apply the mass ratio
\begin{equation}
    \frac{m}{M} \lesssim \frac{1}{10}
\end{equation}
throughout all pairs of bodies in the submoon system.

We may finally define the standard Lagrangian of the $(i-j)$th subsystem as 
\begin{equation}
    \mathcal{L}_{i-j} = T_{i-j}-U_{\text{gravi}, i-j}.
\end{equation}

\subsection{The generalized force}

A preliminary form of the Euler-Lagrange equation reads: 
\begin{equation}\label{eq: ELE preliminary}
    \frac{\mathrm{d}}{\mathrm{d}t} \bigg( \frac{\partial T_{i-j}}{\partial \dot{q}_{k-l}}\bigg) - \frac{\partial T_{i-j}}{\partial q_{k-l}} = S_{k-l}
\end{equation}
\citep[see][chap. 9]{Fliessbach_2020}.  The $q_{k-l}$ are the generalized coordinates. They are either $\psi_l$, which denotes the position of a potentially tidally disrupted packet of mass, or $\varphi_{k-l}$  which denotes the angular position of the tidal disturber:
\begin{equation}
    q_{k-l} \in \{\psi_l, \varphi_{k-l} \hspace{1mm} \mid \hspace{1mm} k,l \in \mathcal{H}\}.
\end{equation}

If $q_{k-l}=\psi_l$, the $k$ index is redundant since the surface angle $\psi_l$ only ever depends on the disturbed body $l$, not the tidal disturber $k$. 

% Soon, we will construct sums over all pairs of tidal disturbers and disturbed bodies, $(i,j)$. 

On the right hand side of Eq. \eqref{eq: ELE preliminary}, we find the "generalized force" component, $S_{k-l}$. It is given by 
\begin{equation}
    S_{k-l}=-\frac{\partial U_{\text{total},i-j}(q,t)}{\partial q_{k-l}}.
\end{equation}

We pull the gravitational part of the total potential energy in $S_{k-l}$ to the left hand side. Because the gravitational energy has a trivial derivative with respect to time, we can isolate the standard Lagrangian $\mathcal{L}_{i-j}$ on the left hand side from the purely dissipative tidal energy $U$ on the right hand side:
\begin{equation}\label{eq: singular ELE}
    \frac{\mathrm{d}}{\mathrm{d}t} \bigg( \frac{\partial \mathcal{L}_{i-j}}{\partial \dot{q}_{k-l}}\bigg) - \frac{\partial \mathcal{L}_{i-j}}{\partial q_{k-l}} = - \frac{\partial U_{\text{tidal},i-j}}{\partial q_{k-l}}
\end{equation}

All that is left is to construct the total Lagrangian on the left hand side and the total tidal energy on the right hand side. 

For the left hand side of \eqref{eq: singular ELE}, one needs to perform a sum over all rotational kinetic energies, as well as a sum over all orbital and gravitational energies of each subsystem. Denoting the set of nearest neighbors (counting each subsystem once) as $\mathbb{P}_1=\{(1,2), (2,3), (3,4), ...\}$: 
%summing once over all tidally disturbed bodies $j$ ensures that all necessary rotational energy contributions are included (see Eq. \eqref{eq: Kinetic Energy}). Performing another sum over the remaining $i$ index and demanding $i=j+1$ incorporates the kinetic energies and pair-interactions due to the mutual gravitational attraction of the bodies, with each pair being counted one time. Therefore, the total standard Lagrangian is 

%\begin{equation}
  %  \mathcal{L}_{\text{all}} = \sum_{i=j+1}\sum_{j} \mathcal{L}_{i-j}.
%\end{equation}
\begin{equation}
    \mathcal{L}_{\text{all}} = \sum_j T_{j}^{\mathrm{{(rot.)}}}+ \sum_{(i,j)\in \mathbb{P}_1} T_{i-j}^{\mathrm{{(orbit.)}}}- U_{\text{gravi}, i-j}
\end{equation}

For the right hand side of \eqref{eq: singular ELE}, we note that while a primary raises a tide on a secondary and vice-versa, the strength of these interactions are not equal, unlike the standard gravitational energy between two bodies. We therefore must perform two sums per subsystem instead of one.
%, to account for the fact that the tidal energy Eq. \eqref{eq: i-j-th Tidal energy} is asymmetric in its indices.
Since as per our assumptions the only bodies that tidally disturb a body $j$ are its nearest neighbors (neglecting secondary tidal effects like submoon-planet interaction), we can perform the sum over the set of adjacent pairs $\mathbb{P}_2$ such that each subsystem is counted twice: 
%\begin{equation}
  %  \mathbb{P}_2=\{(i,j) \hspace{1mm}\mid \hspace{1mm} i,j\in \mathcal{H}, \hspace{1mm} \abs{i-j}=1\}
%\end{equation}
\begin{equation}
     \mathbb{P}_2= \{(1,2), (2,1 ), (2,3 ), (3,2),...\}.
\end{equation}
%Demanding that the distance between $i$ and $j$ is $1$ ensures that $i \neq j$ (a body doesn’t tidally disturb itself) and that bodies that tidally disturb eachother are in direct orbit of eachother.
The total tidal energy is therefore 
\begin{equation}
    U_{\text{all}} = \sum_{(i,j)\in \mathbb{P}_2} U_{\text{tidal},i-j}.
\end{equation}
Finally, we get: 
\begin{equation}
    \frac{\mathrm{d}}{\mathrm{d}t} \frac{\partial \mathcal{L}_{\text{all}}}{\partial \dot{q}_{k-l}} - \frac{\partial \mathcal{L}_{\text{all}}}{\partial q_{k-l}}
    = - \frac{\partial U_{\text{all}}}{\partial q_{k-l}}.
\end{equation}

In the following, we set $n=4$ to extract the differential equations for a submoon system by taking the derivatives with respect to the surface angle $\psi_l$ and the orbit angle $\varphi_{k-l}$.

\section{Differential equations}\label{sec: Differential equations}

The Euler-Lagrange equation for the surface angle coordinate $q_{k-l}=\psi_l$ can be brought into the form of

\begin{equation*}
      I_l \dot{\Omega}_l=-\sum_{(i,j)\in\mathbb{P}_2} \frac{3}{2}Gm_i^2k_{2j}R_j^5Q_j^{-1}\mathrm{sgn}(\Omega_j-n_{i-j})
        \delta_{lj} a^{-6}_{i-j}
\end{equation*}

and for the orbital angular coordinate $q_{k-l}=\varphi_{k-l}$ we arrive at 

\begin{align}
  \frac{1}{2}\frac{m_lm_k}{m_l+m_k} 
    \bigg( 
    \frac{\mu_{l-k}}{a_{l-k}}\nonumber
    \bigg)^{\frac{1}{2}}\dot{a}_{l-k}  &= \frac{3}{2}Gm_k^2R_l^5k_{2l}Q_l^{-1} \cdot \mathrm{sgn}(\Omega_l-n_{k-l})
    a^{-6}_{k-l},
\end{align}

where $\delta_{ij}$ is the Kronecker delta. For a submoon system, all indices $i,j,k,l$ are elements of $\mathcal{H}=\{1,2,3,4\}$ and the sets of adjacent pairs counting each subsystem once and twice are $\mathbb{P}_1=\{(1,2),(2,3),(3,4)\}$ and $\mathbb{P}_2=\{(1,2), (2,1), (2,3), (3,2), (3,4), (4,3)\}$, respectively.  Performing iterations over all elements of these sets yields the differential equations that describe the system\footnote{For a symbolic computation, see \href{https://github.com/iason-saganas/stability-of-submoons/blob/2b12c7ae5d55e432e3ceb5210804ab761379ac4f/media/jupyter_notebooks/ELE-differential-equation-extractor.ipynb}{this jupyter notebook}. }:

\begin{flalign*}
&
    \dot{a}_{\mathrm{m-sm}} = \frac{3R_{\mathrm{m}}^5\mathrm{sgn}(\Omega_{\mathrm{m}}- n_{\mathrm{m-sm}})\sqrt{\mu}_{\mathrm{m-sm}}k_{\mathrm{2m}}m_{\mathrm{sm}}}{Q_{\mathrm{m}}m_{\mathrm{m}}}a_{\mathrm{m-sm}}^{-11/2}
&
\end{flalign*}

\begin{flalign*}
&
    \dot{a}_{\mathrm{p-m}} = \frac{3R_{\mathrm{p}}^5\mathrm{sgn}(\Omega_{\mathrm{p}}-n_{\mathrm{p-m}})\sqrt{\mu}_{\mathrm{p-m}}k_{\mathrm{2p}}m_{\mathrm{m}}}{Q_{\mathrm{p}}m_{\mathrm{p}}}a_{\mathrm{p-m}}^{-11/2}
&
\end{flalign*}

\begin{flalign*}
&
    \dot{a}_{\mathrm{s-p}} = \frac{3R_{\mathrm{s}}^5\mathrm{sgn}(\Omega_{\mathrm{s}}-n_{\mathrm{s-p}})\sqrt{\mu}_{\mathrm{s-p}}k_{\mathrm{2s}}m_{\mathrm{p}}}{Q_{\mathrm{s}}m_{\mathrm{s}}}a_{\mathrm{s-p}}^{-11/2}
&
\end{flalign*}
\begin{flalign*}
&
    \dot{\Omega}_{\mathrm{m}} = -\frac{3GR_{\mathrm{m}}^5k_{\mathrm{2m}}\mathrm{sgn}(\Omega_{\mathrm{m}}-n_{\mathrm{p-m}})m_{\mathrm{p}}^2}{2I_{\mathrm{m}}Q_{\mathrm{m}}}a_{\mathrm{p-m}}^{-6} \\ & \hspace{10mm} - \frac{3GR_{\mathrm{m}}^5k_{\mathrm{2m}}\mathrm{sgn}(\Omega_{\mathrm{m}}-n_{\mathrm{m-sm}})m_{\mathrm{sm}}^2}{2I_{\mathrm{m}}Q_{\mathrm{m}}}a_{\mathrm{m-sm}}^{-6}
&
\end{flalign*}

\begin{flalign*}
&
    \dot{\Omega}_{\mathrm{p}} = -\frac{3GR_{\mathrm{p}}k_{\mathrm{2p}}\mathrm{sgn}(\Omega_{\mathrm{p}}-n_{\mathrm{s-p}})m_{\mathrm{s}}^2}{2I_{\mathrm{p}}Q_{\mathrm{p}}}a_{\mathrm{s-p}}^{-6} \\ & \hspace{10mm} - \frac{3GR_{\mathrm{p}}k_{\mathrm{2p}}\mathrm{sgn}(\Omega_{\mathrm{p}}-n_{\mathrm{p-m}})m_m^2}{2I_{\mathrm{p}}Q_{\mathrm{p}}}a_{\mathrm{p-m}}^{-6}
&
\end{flalign*}
\begin{flalign*}
&
    \dot{\Omega}_{\mathrm{s}} = -\frac{3GR_{\mathrm{s}}^5\mathrm{sgn}(\Omega_{\mathrm{s}}-n_{\mathrm{p-s}})k_{\mathrm{2s}}m_{\mathrm{p}}^2}{2I_{\mathrm{s}}Q_{\mathrm{s}}}a_{\mathrm{s-p}}^{-6}.
    &
\end{flalign*}

\begin{figure}
    \centering
    \includegraphics[width=0.9\linewidth]{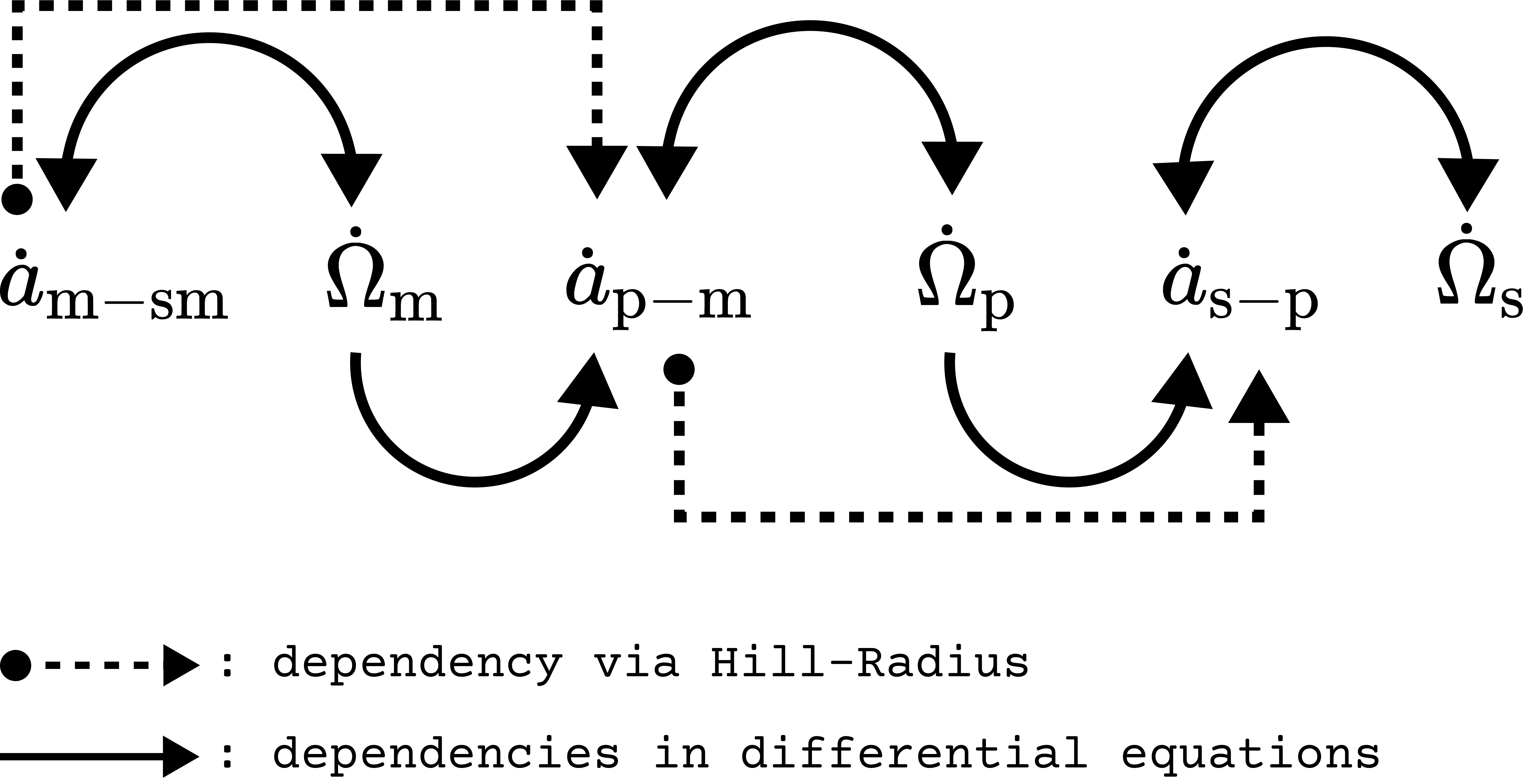}
    \caption{Network of inter dependencies of the variables of interest. The variables are not only coupled through the differential equations, but also through the Hill-Radius, which is computed in order to calculate the upper escape bound, the critical semi-major axis, $a_{\text{crit}}$. Since the semi-major axes change in time, a numerical solver needs to calculate the upper bound $a_{\text{crit}}$ dynamically in each time step in order to not trigger premature termination.}
    \label{fig: Dependency network}
\end{figure}

A sketch of which variables are interdependent in these differential equations is given in Fig. \ref{fig: Dependency network}

\begin{figure}
    \centering
    \begin{subfigure}{0.48\textwidth}
        \centering
        \includegraphics[width=\textwidth]{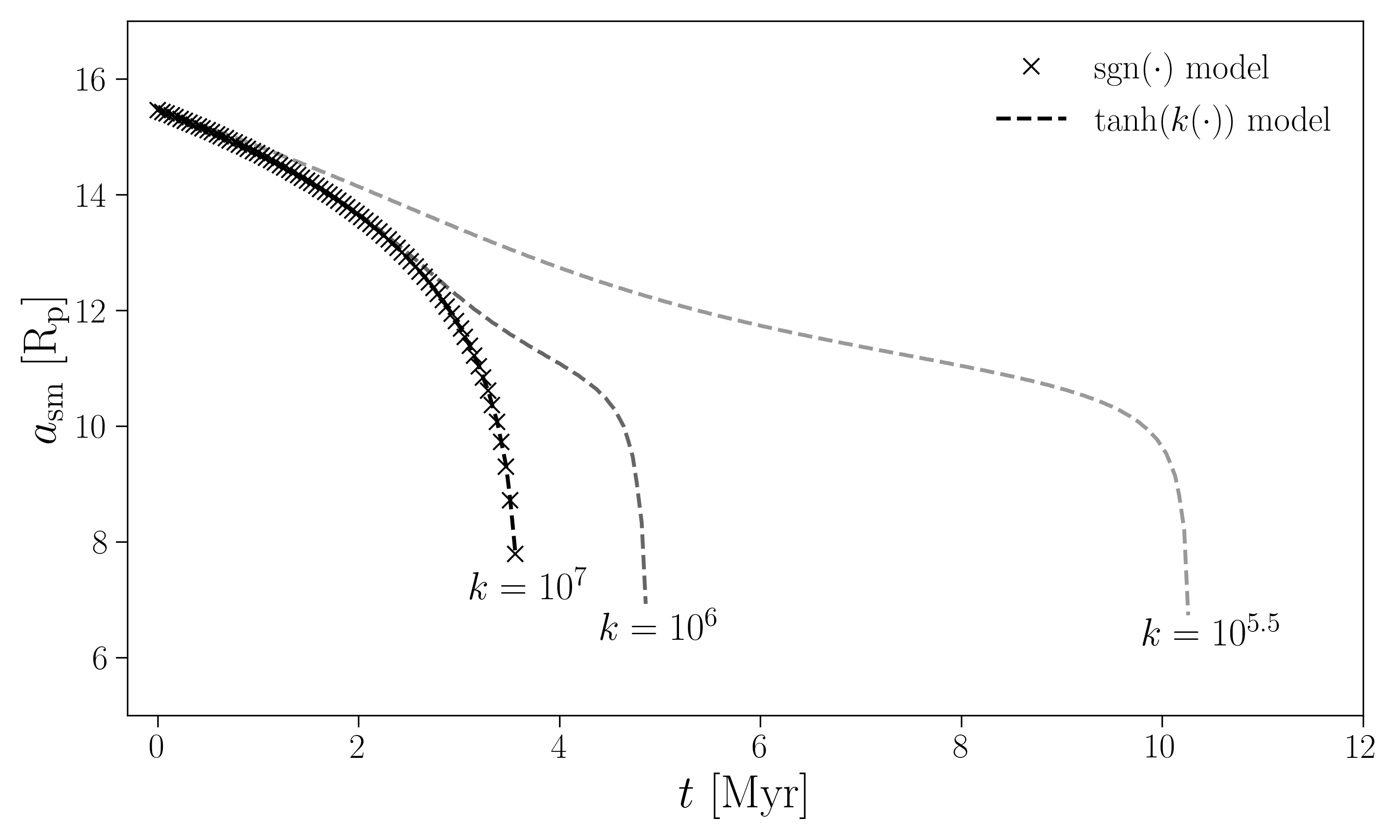}
        \caption{Effect of changing the $k$ parameter in the $\mathrm{tanh}(k(\cdot))$ model on the evolution of the state-vector. As an example, the evolution of the submoon semi-major axis in Earth-radii is shown. All $k$ parameters produce the same outcome for the submoon (disintegration at the Roche-limit), but the time of termination differs by some $\mathrm{Myr}$. 
        %The value $k=10^6$ is taken to be satisfactory to balance precision with speed.
        }
        \label{fig: subfig_a effect of different k parameters}
    \end{subfigure}

    \begin{subfigure}{0.48\textwidth}
        \centering
        \includegraphics[width=\textwidth]{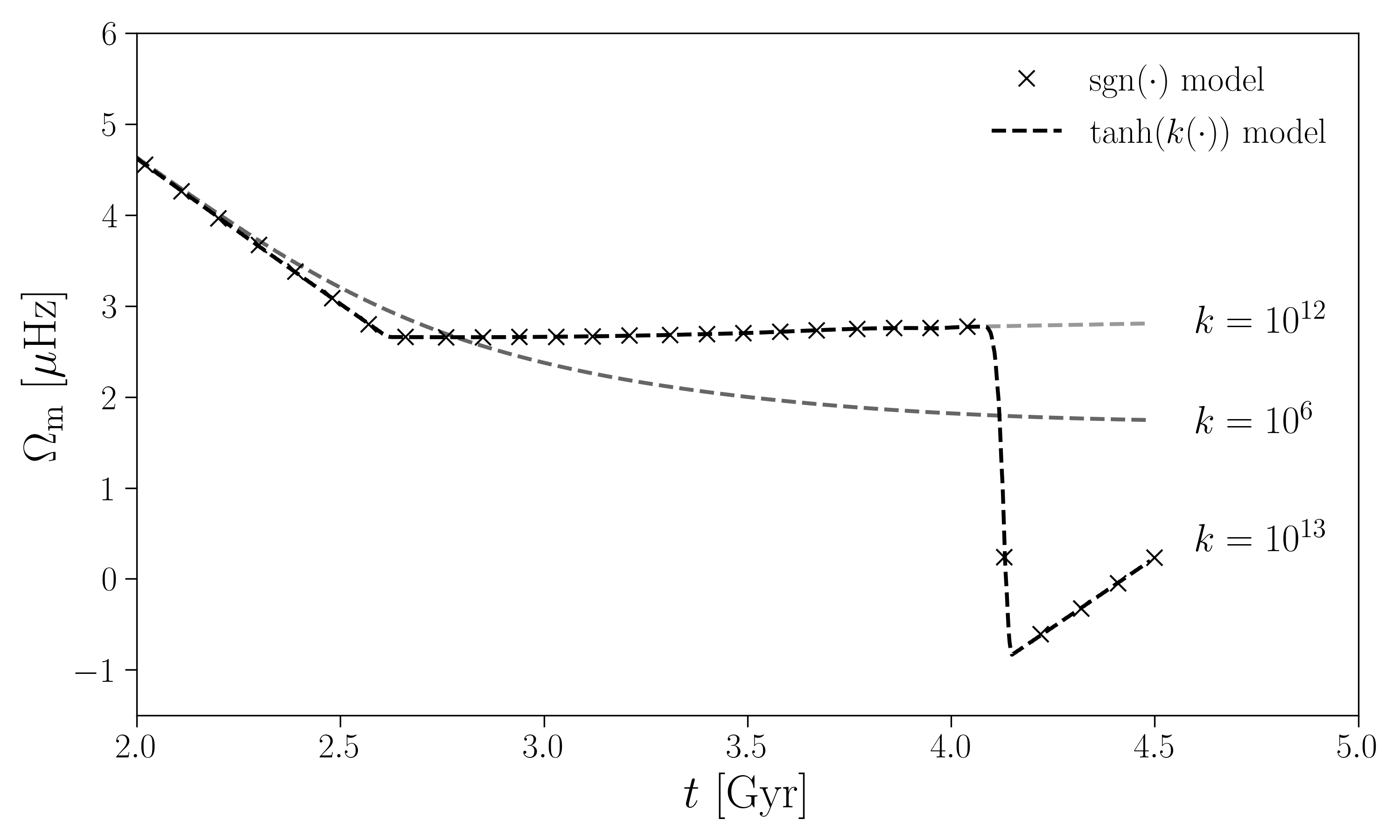}
        \caption{One particular example in which the shape of the evolution in the moon's spin frequency is not well captured by $k$-values smaller than $k=10^{13}$.}
        \label{fig: subfig_b particularly unsmooth evolution}
    \end{subfigure}

    \begin{subfigure}{0.48\textwidth}
        \centering
        \includegraphics[width=\textwidth]{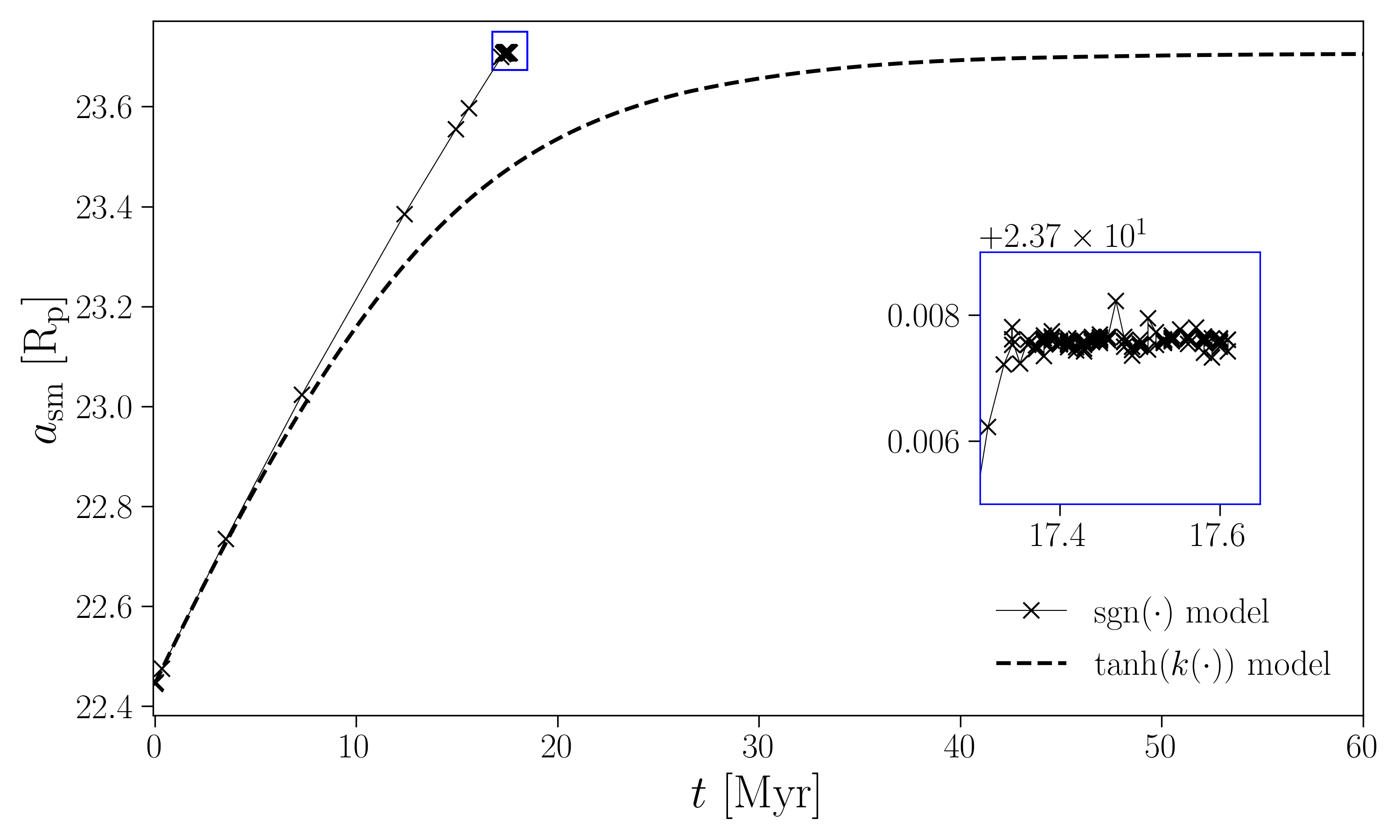}
        \caption{The $\mathrm{sgn}(\cdot)$ model often exhibits numerical issues when there are steep gradients in the solution, in terms of small oscillations on time-scales of $10^4-10^5\mathrm{yr}$ (inset plot). The $\mathrm{tanh}(k(\cdot))$ solution shows a smoothing out of such features. In this case, the $\mathrm{sgn}(\cdot)$ model appears to oscillate around a line of the same height at which the $\mathrm{tanh}(k(\cdot))$ model stabilizes later on.} 
        \label{fig: subfig_c small scale oscillations}
    \end{subfigure}

    \caption{(a) Effect of changing the $k$ parameter in the $\mathrm{tanh}(k(\cdot))$ model,  (b) Particularly unsmooth evolution exhibited by the $\mathrm{sgn}(\cdot)$ model and (c) small scale oscillations in the $\mathrm{sgn}(\cdot)$ model.}
    \label{fig: three_row_halfpage}
\end{figure}

\section{Numerical results}\label{sec: Numerical Results}

\subsection{Numerical solver and stability considerations}

In the following, we will use a numerical integrator to solve the differential equations presented in the previous section for two kinds of star-planet-moon-submoon systems: One that represents our Earth system in regard of the masses of the bodies, and one that represents the Kepler1625 system. In the analysis, we will fix the initial values of $\Omega_{\text{s}}^{(0)}$, $\Omega_{\text{p}}^{(0)}$ and $\Omega_{\text{m}}^{(0)}$ to some self-chosen, sensible values, while iterating through all combinations of semi-major axes initial values, $(a_{\text{sm-m}}^{(0)}, a_{\text{m-p}}^{(0)}, a_{\text{p-s}}^{(0)})$. The ranges of the latter are constrained by their respective Roche-limits and critical semi-major axes. The stable 3D parameter space spanned by $(a_{\text{sm-m}}, a_{\text{m-p}}, a_{\text{p-s}})$ can then be visualized by marking the points that:

\begin{itemize}
    \item  represent initial states $(a_{\text{sm-m}}^{(0)}, a_{\text{m-p}}^{(0)}, a_{\text{p-s}}^{(0)})$ for which the solver successfully reached the end of the integration interval ($4.5 \mathrm{Gyr}$),
    \item or that represent the evolved, final states $(a_{\text{sm-m}}^{(f)}, a_{\text{m-p}}^{(f)}, a_{\text{p-s}}^{(f)})$ of the aforementioned stable initial states.
\end{itemize}

The \href{https://github.com/iason-saganas/stability-of-submoons/blob/2b12c7ae5d55e432e3ceb5210804ab761379ac4f/differential_equation_solver.py}{solver} employed in this work is based on \texttt{SciPy}'s \texttt{solve\textunderscore ivp} routine with the method set to \texttt{Radau}. The \texttt{Radau} solver is an implicit Runge-Kutta method, meaning that in each new time step $t_{n+1}$, the function value there, $y_{n+1}$, is found by solving an equation that depends on both $y_{n+1}$ itself and an average of derivatives over the interval $[t_{n},t_{n+1}]$.

Importantly, implicit methods like \texttt{Radau} have the property of being "A-Stable", which aids in obtaining meaningful numerical results due to the fact that the stability region covers the entire left half plane of $\mathbb{C}$. This is especially important when dealing with stiff differential equations. These typically describe systems consisting of two or more dynamic components evolving over different characteristic time-scales, often encountered in, e.g., chemistry \citep{RadauSolver}.  

Indeed, the differential equations studied in this work contain one crucial component in this regard: The factor $\mathrm{sgn}(\Omega -n)$, which gives the overall sign of the change of rate depending on the relative tidal forcing frequency. The acting torque $\Gamma$, derived as the gradient of the non-central potential, can be expressed as 
\begin{equation}
    \Gamma := \Gamma_0 \mathrm{sgn}(\Omega -n)
\end{equation}
(see \cite{Murray_Dermott_2000}, Eq. 4.159), with some constant $\Gamma_0>0$.

While the evolution of a system governed by such torques might be smooth in general, once any tidal forcing frequency $\Omega -n$ reaches and crosses $0$, the acting torque will exhibit a non-continuous jump between its extreme amplitudes, e.g., from $+\Gamma_0$ to $-\Gamma_0$.

This creates an unrealistic physical scenario: The location of the tidal bulge jumps from $+2\delta_{\text{lag}}$ to $-2\delta_{\text{lag}}$ in an instant. This is not only unphysical, but introduces numerical issues in the solutions in the form of small-scale oscillations that make the integration prohibitively computationally expensive, see inset plot of Fig. \ref{fig: subfig_c small scale oscillations}. 

We propose a smooth parameterization of this "jump" by replacing the $\mathrm{sgn}(\cdot)$ function with the hyperbolic tangent function in the differential equations governing the evolution:
\begin{equation}
    \mathrm{sgn}(\Omega -n ) \longrightarrow \mathrm{tanh}(k(\Omega -n)), 
\end{equation}
with an arbitrary width parameter $k$. Henceforth, we will refer to the original differential equations, including the sudden, non-continuous jump, as the "$\mathrm{sgn}(\cdot)$ model" and its smooth parameterization via the hyperbolic tangent as the "$\mathrm{tanh}(k(\cdot))$ model".

The latter effectively models a time-dependent quality factor $Q$, leading to a more realistic torque, which we found crucial for increasing the numerical stability. 

Our approach is the following: Each time stiffness, i.e. small-scale oscillations, are detected in the solutions at a time point $t_{\text{cut}}$, the integration is terminated and the iteration re-run with the more numerically robust $\mathrm{tanh}(k(\cdot))$ model. 

If the $\mathrm{tanh}(k(\cdot))$ solution agrees with the $\mathrm{sgn}(\cdot)$ solution up to the point $t_{\mathrm{cut}}$ within $1\%$ at all time steps, we assume the $\mathrm{tanh}(k(\cdot))$ approximation to be satisfactory for all time steps $t>t_{\text{cut}}$ as well. 

If there are deviations larger than $1\%$ early on for $t<t_{\mathrm{cut}}$ we inspect the deviation by eye. If the general trend suggested by the $\mathrm{sgn}(\cdot)$ model for $t<t_{\text{cut}}$ is still captured by the $\mathrm{tanh}(k(\cdot))$ model, we accept the iteration in spite of deviations larger than $1\%$ (see, e.g., Fig. \ref{fig: subfig_c small scale oscillations}). 

For the parameter ranges considered, we found that about 20\% of iterations turn out stiff.
We can use the stable, non-stiff iterations, where the $\mathrm{sgn}(\cdot)$ model does reach the end of the integration interval, for a comparison of the two models' numerical solutions. 

For example, in the test case "A" of Table \ref{table: Most stable omega_i initial condition}, the $\mathrm{sgn}(\cdot)$ model produced non-stiff solutions for $192$ sets of initial parameters. Recomputing these initial value problems with the $\tanh(k(\cdot))$ model (setting $k=10^6$) and comparing the average relative difference in the final submoon semi-major axis, i.e.
\begin{equation*}
    \texttt{res} = \bigg\langle \frac{a_{\mathrm{sm, sgn(\cdot)}}^{(\text{final})}-a_{\mathrm{sm, tanh}(k(\cdot))}^{(\text{final})}}{a_{\mathrm{sm, sgn(\cdot)}}^{(\text{final})}} \bigg\rangle,
\end{equation*}
one finds a residual of $\texttt{res}\approx 0.1\%$. Figure \ref{fig: subfig_a effect of different k parameters} shows the solution with the largest difference between the two models in the final submoon semi-major axis (relative difference of about $11\%$).
There, we also show evolutions of the $\mathrm{tanh}(k(\cdot))$ model under different $k$ parameters. We found that 
\begin{equation}
    k=10^6
\end{equation}
strikes a balance between accuracy and computational affordability. The run with the next-largest relative deviation of about $3\%$ is shown in Fig. \ref{fig: subfig_b particularly unsmooth evolution}. There, we see a particularly unsmooth evolution using the $\mathrm{sgn}(\cdot)$ model, which is only well captured by its $\mathrm{tanh}(k(\cdot))$ counterpart when setting $k\geq 10^{13}$. At that point, the $\mathrm{tanh}(k(\cdot))$ model exhibits the same numerical issues as the differential equations solved with the $\mathrm{sgn}(\cdot)$ function: Rapid oscillations on the time-scale of tens to hundreds of thousands of years that often arise when the solution exhibits "nicks" or sharp corners. We found that choosing a value of $k\sim 10^6$ smoothes such features out while often continuing the trend sketched by the $\mathrm{sgn}(\cdot)$ solution (see Fig. \ref{fig: subfig_c small scale oscillations}).

To further build trust in the numerical solutions, we check that the total angular momentum in each iteration is conserved. Additionally, provided that the signs of the right hand sides of the differential equations remain constant throughout the evolution, the differential equations can be solved analytically. We then compare our numerical solutions to the expected analytical solutions and report any deviations larger than $1\%$. 

A given iteration will terminate if any of the bodies semi-major axes surpasses their respective critical semi-major axis or falls under their respective Roche-Limits.

In Table \ref{table: Most stable omega_i initial condition} we can see the effect of the chosen $(\Omega_i^{(0)})$ initial conditions on the stability of a submoon system with a small test mass of $m_{\mathrm{sm}}=1\% M_{\text{Luna}}$. There is only a slight difference on the total number of iterations having reached a stable lifetime of $4.5\mathrm{Gyr}$ between different $(\Omega_i^{(0)})$ initial conditions. We assume that this test run with $m_{\mathrm{sm}}=1\% M_{\text{Luna}}$ is somewhat representative for other submoon masses as well and continue the further analysis by choosing case A as the initial $(\Omega_i^{(0)})$ condition, meaning the star spins slower around its own axis than the planet does, which in turn takes longer to complete one rotation than the moon.

\begin{table}[!h]
\footnotesize
\tabcolsep=4pt
\processtable{Fraction of states that ended up surviving $4.5\mathrm{Gyr}$ shown for different initial spin frequency conditions $(\Omega_{\mathrm{s}}^{(0)}, \Omega_{\mathrm{p}}^{(0)}, \Omega_{\mathrm{m}}^{(0)})$. The system considered here is made up of a solar-mass star, an Earth-mass planet and a Luna-mass moon with a small test mass of $1\% M_{\text{Luna}}$ for the submoon. 
%The planet's semi-major axis ranged from its Roche-Limit to $10\mathrm{AU}$, thus also fixing ranges of moon and submoon via their Roche-Limits and critical semi-major axes. The total explored parameter-space $(a_{\mathrm{p}}, a_{\mathrm{m}}, a_{\mathrm{sm}})$ spans a $10\times 10\times 10$ large cube. ``Stable iterations'' refers to how many points in the $10\times 10\times 10$ parameter space reached a stable lifetime of $4.5\mathrm{Gyr}$. 
\label{table: Most stable omega_i initial condition}}{
\begin{tabularx}{\columnwidth}{lYYYr}
\rowcolor{Theadcolor}
\textbf{Case} & \multicolumn{3}{c}{\textbf{Initial Rotation Periods}} & \textbf{Stable Iterations} \\
\rowcolor{Theadcolor}
 & \textbf{$P_{\mathrm{s}}^{(0)}$} & \textbf{$P_{\mathrm{p}}^{(0)}$} & \textbf{$P_{\mathrm{m}}^{(0)}$} & \\
A & $20\mathrm{d}$ & $50\mathrm{h}$ & $24\mathrm{h}$ & $23.1\%$ \\
B & $20\mathrm{d}$ & $50\mathrm{d}$ & $10\mathrm{d}$ & $20.8\%$ \\
C & $20\mathrm{d}$ & $24\mathrm{h}$ & $10\mathrm{d}$ & $20.8\%$ \\
D & $5\mathrm{d}$  & $10\mathrm{d}$ & $30\mathrm{d}$ & $20.6\%$ \\
E & $20\mathrm{d}$ & $50\mathrm{d}$ & $40\mathrm{d}$ & $20.6\%$ \\
F & $20\mathrm{d}$ & $10\mathrm{h}$ & $50\mathrm{d}$ & $20.6\%$ \\
\end{tabularx}}{}
\end{table}

\subsection{Earth-like submoon system}\label{subsec: Earth-like submoon-system}

In this section, we analyze a star-planet-moon system with Sun-like, Earth-like, and Luna-like masses, respectively.

For the planet, $(k_2,Q)=(0.3, 280)$ is assumed \citep{Lainey_2016}. For the moon, we set $(k_2, Q)=(0.25, 100)$ (see caption of Fig. 1 of \cite{Kollmeier_2018}). As a rough guess, we apply the same parameters for the submoon. For the star, we employ a rotation-period-dependent estimation of the ratio $Q/k_2$. From \cite{Barker_2022_Q_star}: 
\begin{equation}
    \frac{3}{2}\bigg(\frac{Q}{k_2}\bigg)^{\ast} \approx 10^7 \bigg(\frac{P^{\ast}_{\text{rot}}}{10\hspace{1mm}\mathrm{days}}\bigg)^2,
\end{equation}
where the asterisk refers to the star.

\begin{figure*}
    \centering
    \begin{tabular}{@{\hspace{0.03\textwidth}}c@{\hspace{0.06\textwidth}}c}
        \subcaptionbox{\centering Initial parameter space with submoon \\ mass $m_{\mathrm{sm}}=10^{15}\mathrm{kg}$. \label{subfig: Earth-like-results a}}
        {\includegraphics[height=0.3\textheight]{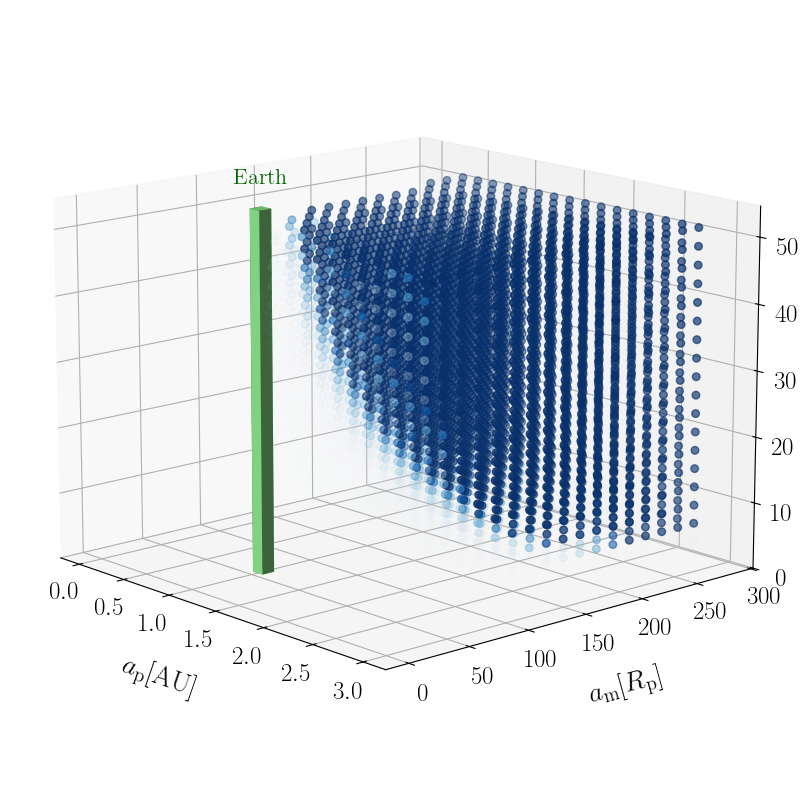}} &

        \subcaptionbox{\centering Parameter space after tidal migration \\ with submoon mass $m_{\mathrm{sm}}=10^{15}\mathrm{kg}$. \label{subfig: Earth-like-results b}}
        {\includegraphics[height=0.3\textheight]{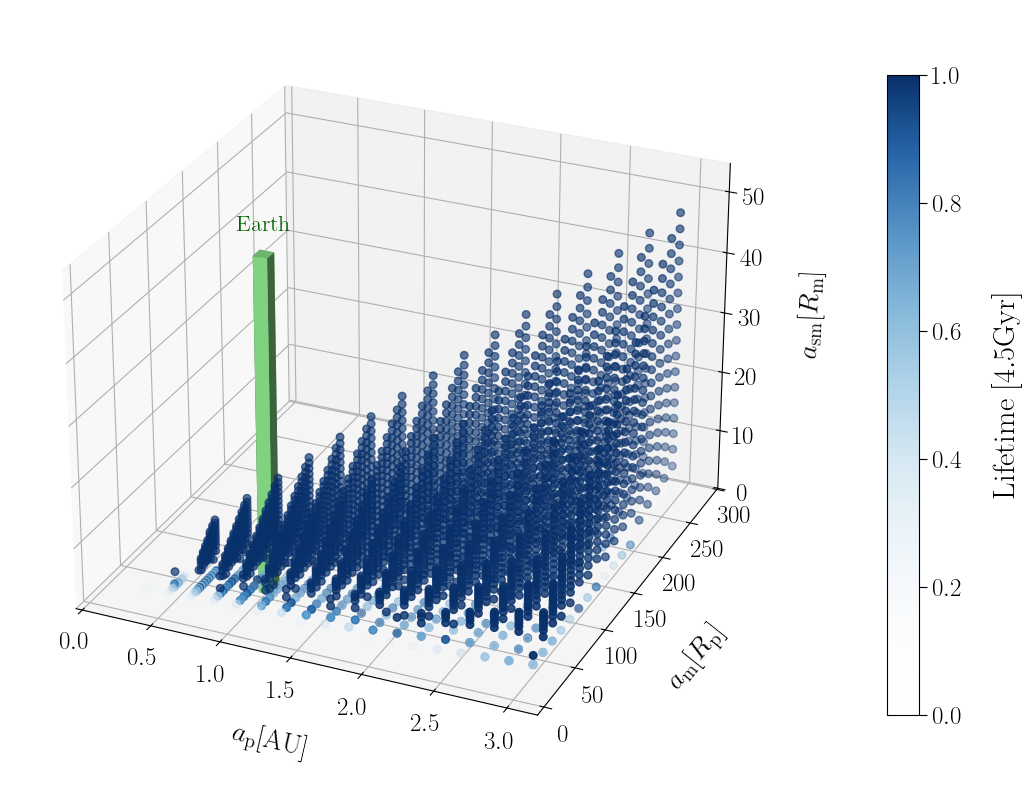}} \\

        \subcaptionbox{\centering Initial parameter space with submoon \\ mass $m_{\mathrm{sm}}=4.6\cdot 10^{17}\mathrm{kg}$. \label{subfig: Earth-like-results c}}
        {\includegraphics[height=0.3\textheight]{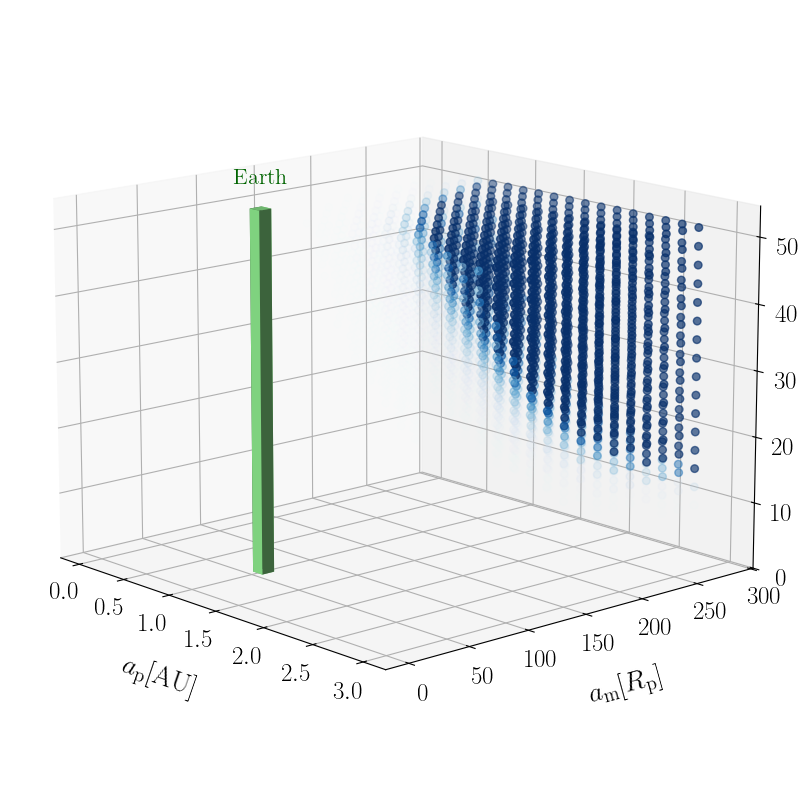}} &

        \subcaptionbox{\centering Parameter space after tidal migration \\ with submoon mass $m_{\mathrm{sm}}=4.6\cdot 10^{17}\mathrm{kg}$. \label{subfig: Earth-like-results d}}
        {\includegraphics[height=0.3\textheight]{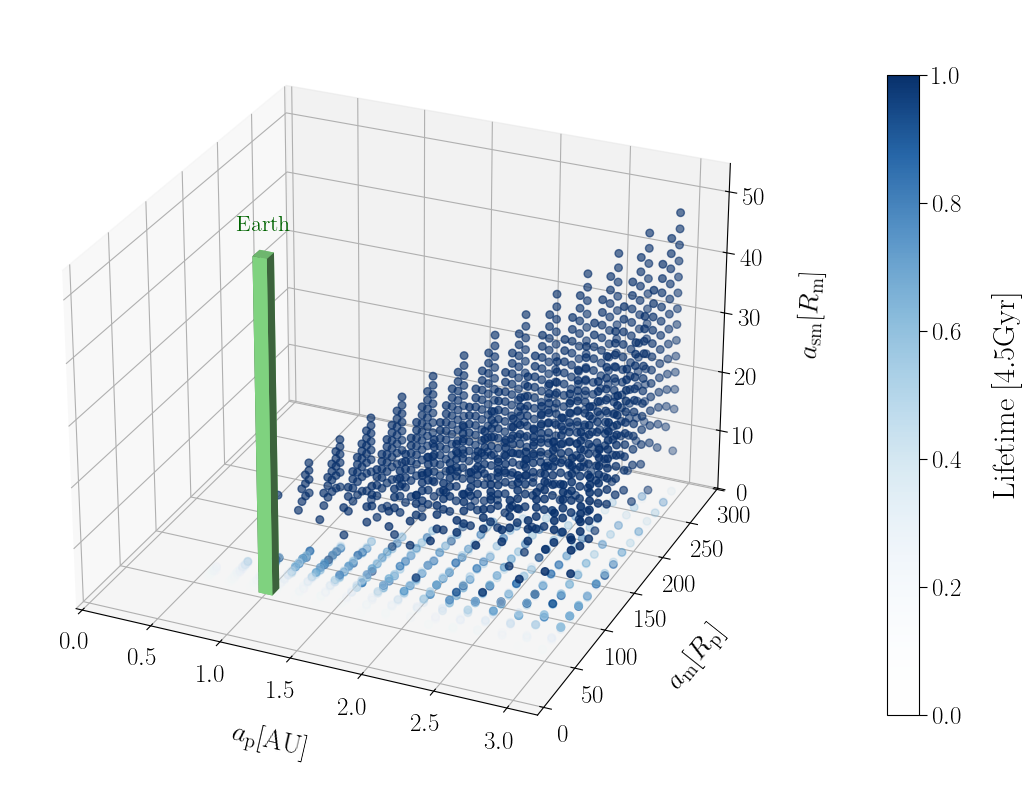}}
    \end{tabular}

    \caption{Stability regions of an Earth-like submoon system. Shown are point clouds in the semi-major axes parameter space which are color-coded according to how long a system described by each of these points is stable before tides act to remove or destroy the moon, submoon or planet. The two rows describe systems of different submoon masses ($10^{15}\mathrm{kg}$ for upper and $4.6\cdot 10^{17}\mathrm{kg}$ for lower). The left columns show initial states, the right columns the tidally evolved initial states. In each plot, the green pillar marks the approximate configuration of our current Earth-system in this parameter space (with an unconstrained $z$-axis since Earth harbors no natural submoon). The units on the $y-$ and $z-$axes are Earth-radii and Luna-radii respectively.}
    \label{fig: Earth-like-results main}
\end{figure*}

To begin, we employ a $10^{15}\mathrm{kg}$ submoon mass, which corresponds to the $5\mathrm{km}$ critical line of \cite{Kollmeier_2018} (dotted line in Fig. 1 therein).The resulting stability region in the initial semi-major axis space is seen in Fig. \ref{subfig: Earth-like-results a}, where we see a transition surface to the upper right of which points are stable. 

In the panel to the right, Fig. \ref{subfig: Earth-like-results b}, the locations of the same points after having undergone tidal migration are marked: We see that the current configuration of our Earth-system (green pillar) intersects with the stable point cloud, indicating that Earth in its contemporary configuration could have hosted a small submoon, as \cite{Kollmeier_2018} have already found. 

In general, we make the observation that stable points in the semi-major axis parameter space represent planets, moons and submoons on wide orbits which then migrate inwards by varying degrees. The wider the planet and moon orbit from their hosting bodies, the nearer the submoon may be to the moon and still be stable, as seen by the curvature of the transition surface in Fig. \ref{subfig: Earth-like-results a}.

By choosing a coarser grid of the searched parameter space than in Fig. \ref{subfig: Earth-like-results a} and \ref{subfig: Earth-like-results b}, we may gauge how large a submoon could have been and still be hosted in an Earth-like system by rerunning the analysis and sequentially increasing the submoon mass. Indeed, there are only a few, far-out points in the parameter space that are stable for a submoon on the order of $4.6\cdot 10^{17}\mathrm{kg}$ (see Fig. \ref{subfig: Earth-like-results c}). Some of these points do tidally migrate inwards toward Earth's current configuration, but don't arrive there exactly (see Fig. \ref{subfig: Earth-like-results d}). In the resolution used in this run, the nearest island of stability similar to Earth's current configuration ($(a_{\mathrm{p}}=1\mathrm{AU}, a_{\mathrm{m}}=60R_{\mathrm{p}})$) is around values $(a_{\mathrm{p}}\sim 1.1\mathrm{AU}, a_{\mathrm{m}}\sim 80R_{\mathrm{p}})$.

To conclude, we numerically refine the orbital constraints of the previously suggested upper bound on a submoon's mass for the Earth system ($m_{\mathrm{sm}}^{\text{(max)}}\sim 4.6\cdot 10^{17}\mathrm{kg}$ in \cite{Kollmeier_2018}) to lie somewhat outside Earth's contemporary configuration. Rather, a mass of $10^{15}\mathrm{kg}$ is much more likely to have remained in a stable orbit within Earth’s present-day configuration.

\subsection{Warm-Jupiter-like submoon system}\label{subsec: Warm-Jupiter-like submoon system}

\begin{figure*}
    \centering
    \begin{tabular}{@{\hspace{0.03\textwidth}}c@{\hspace{0.06\textwidth}}c}
        \subcaptionbox{\centering Initial parameter space with submoon \\ mass $m_{\mathrm{sm}}=10^{-1}M_{\text{Luna}}$. \label{subfig: Jup-like-results a}}
        {\includegraphics[height=0.3\textheight]{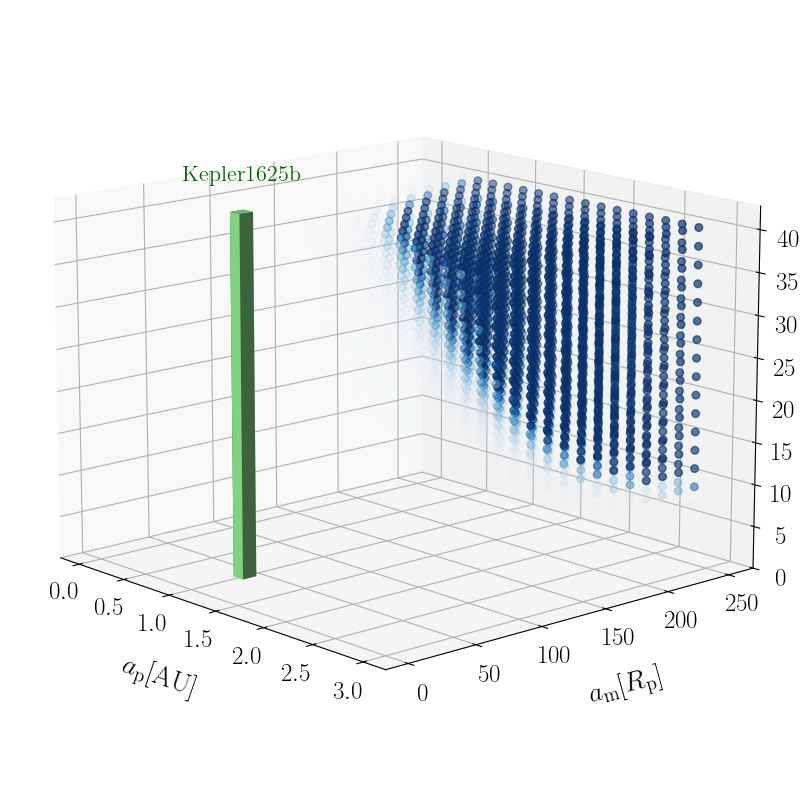}} &

        \subcaptionbox{\centering Parameter space after tidal migration with \\ submoon mass $m_{\mathrm{sm}}=10^{-1}M_{\text{Luna}}$. \label{subfig: Jup-like-results b}}
        {\includegraphics[height=0.3\textheight]{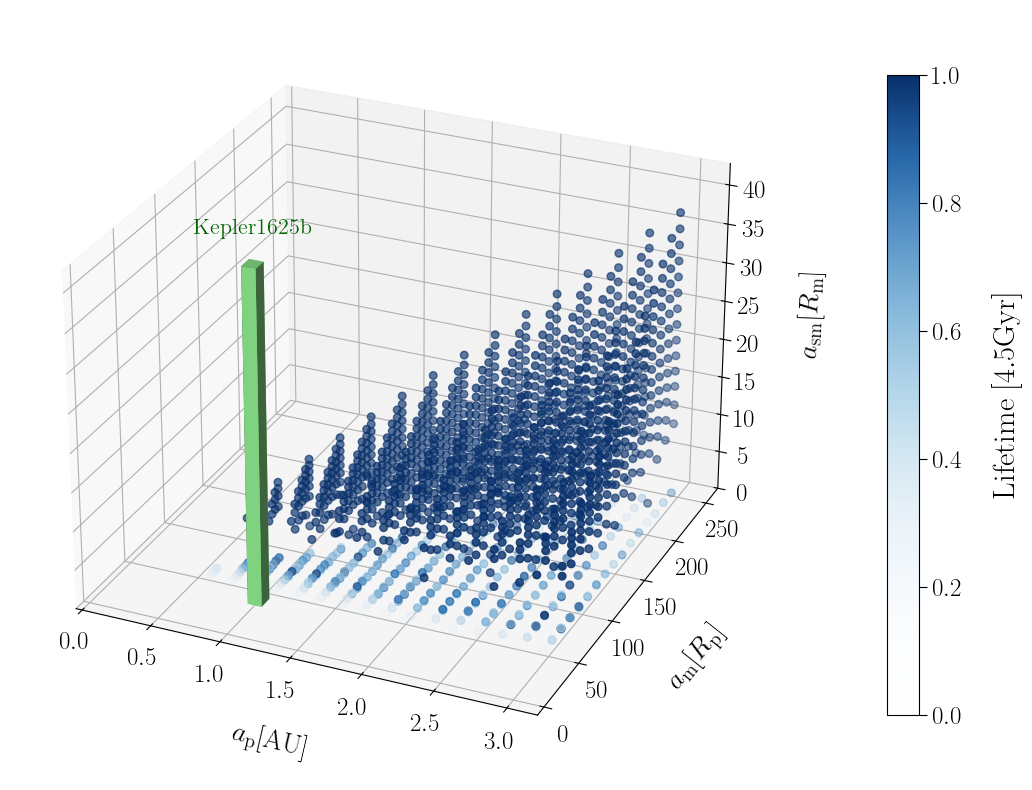}} \\

        \subcaptionbox{\centering Initial parameter space with submoon \\ mass $m_{\mathrm{sm}}=1.8M_{\oplus}$. \label{subfig: Jup-like-results c}}
        {\includegraphics[height=0.3\textheight]{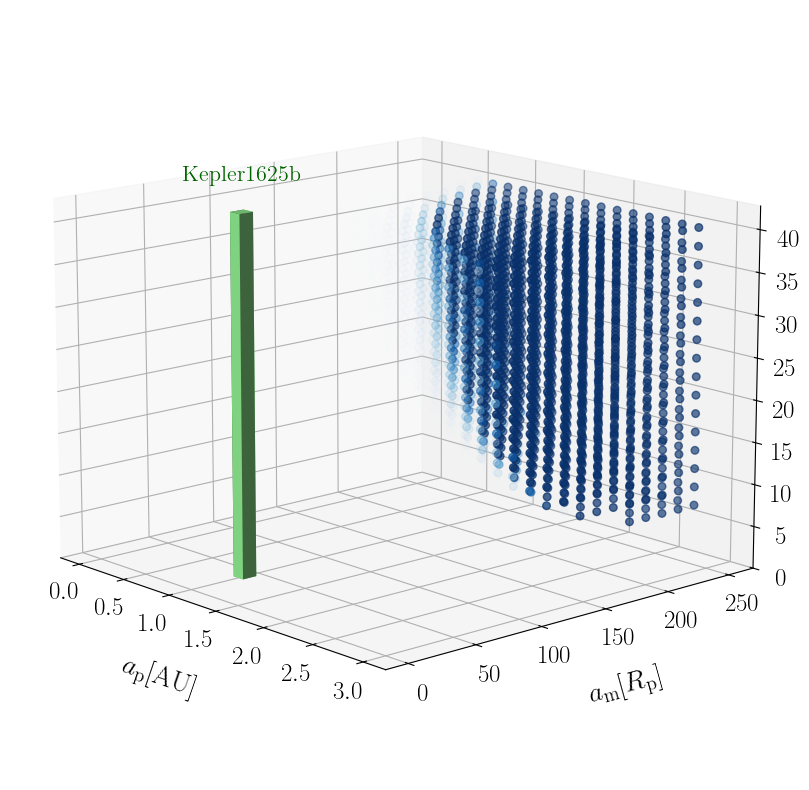}} &

        \subcaptionbox{\centering Parameter space after tidal migration with \\ submoon mass $m_{\mathrm{sm}}=1.8M_{\oplus}$. \label{subfig: Jup-like-results d}}
        {\includegraphics[height=0.3\textheight]{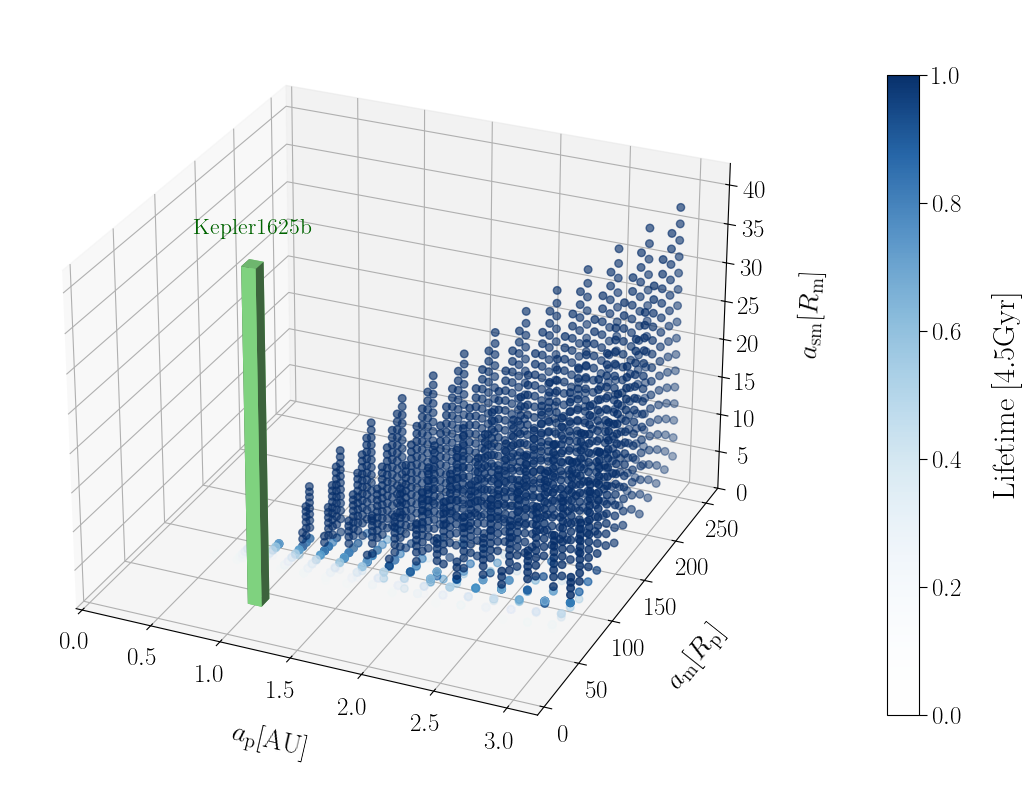}}
    \end{tabular}

    \caption{Stability regions of a Warm-Jupiter-like submoon system. Shown are point clouds in the semi-major axes parameter space which are color-coded according to how long a system described by each of these points is stable before tides act to remove or destroy the moon, submoon or planet. The two rows describe systems of different submoon masses ($m_{\mathrm{sm}}=10^{-1}M_{\text{Luna}}$ for upper and $m_{\mathrm{sm}}=1.8M_{\oplus}$ for lower). The left columns show initial states, the right columns the tidally evolved initial states. In each plot, the green pillar approximately marks the current configuration Kepler1625b in this parameter space (with an unconstrained $z$-axis since there is no evidence to the effect of an existing submoon). The units on the $y-$ and $z-$ axes are eleven Earth-radii and four Earth-radii respectively.}
    \label{fig: Warm-Jupiter-like-results main}
\end{figure*}

In this section, with reference to Kepler1625, we analyze a system consisting of a solar-mass star, that hosts a $3M_{\mathrm{J}}$ mass planet \citep{KeplerMass} that in turn hosts a $18.6 M_{\oplus}$ moon (see Table 2 of \cite{Kipping_exomoon}). For the planet, we assume a Jupiter-like $k_2/Q$ ratio of $1.102\cdot 10^{-5}$ \citep{Lainey_2016}. We model the moon as Neptune-like, i.e. we set $k_2=0.127$ and $Q\approx 5000$ (see \cite{GAVRILOV1977443} Table 1 and last paragraph respectively, assuming $Q_{\text{Neptune}}\approx Q_{\text{Uranus}}$). The star is modeled in the same way as described in the section on Earth-like submoon systems. If the submoon has a mass $m_{sm} \leq 0.5$, it is described by the submoon parameters from the previous section, whereas if it exceeds that threshold, it is modeled as the Earth-like object discussed there.

The maximum submoon mass found by \cite{Kollmeier_2018} for the Kepler1625 system under moon-submoon tides was on the order of $\sim 10^{-3}M_{\text{Luna}}$. 

We find that for a submoon mass of $10^{-1}M_{\text{Luna}}$, although rare, there exist points in the initial state space that lead to systems approximating the current configuration of Kepler1625b with $a_{\mathrm{p}}\sim 1\mathrm{AU}$ and $a_{\mathrm{m}}\sim 67 R_{\mathrm{p}}$ as the nearest such points (compare against configuration of Kepler1625b at $a_{\mathrm{p}}=1\mathrm{AU}$ and $a_{\mathrm{m}}=40R_{\mathrm{p}}$, green pillar in Fig. \ref{subfig: Jup-like-results b}).

The ability of the system to host a massive submoon of around $1.8M_{\oplus}$ improves significantly when the separation of the Neptune-like moon to its Jupiter-like primary is $\gtrsim 100 R_{\mathrm{p}}$ (see Fig. \ref{subfig: Earth-like-results d}). Note that a submoon mass of $1.8M_{\oplus}$ corresponds to our chosen maximum mass threshold of $10\%$ of the moon’s mass.

If tectonic activity and a long-lived atmosphere are base prerequisites for life, it might be possible that a less massive submoon of around $0.5M_{\oplus}$ \citep{Raymond_2007} might suffice in this context instead of the $1.8M_{\oplus}$ mass.

Interestingly, the total fraction of stable iterations decreases from $18.1\%$ ($1.8{M_{\oplus}}$) to $10.2\%$ when considering the smaller $0.5M_{\oplus}$ submoon mass. The large fraction of this difference is due to a higher disintegration rate of the submoon at its Roche-limit ($4559$ vs. $5226$ destruction events for the higher- and lower-mass submoon respectively).

\begin{figure}
    \centering
    \includegraphics[width=0.9\linewidth]{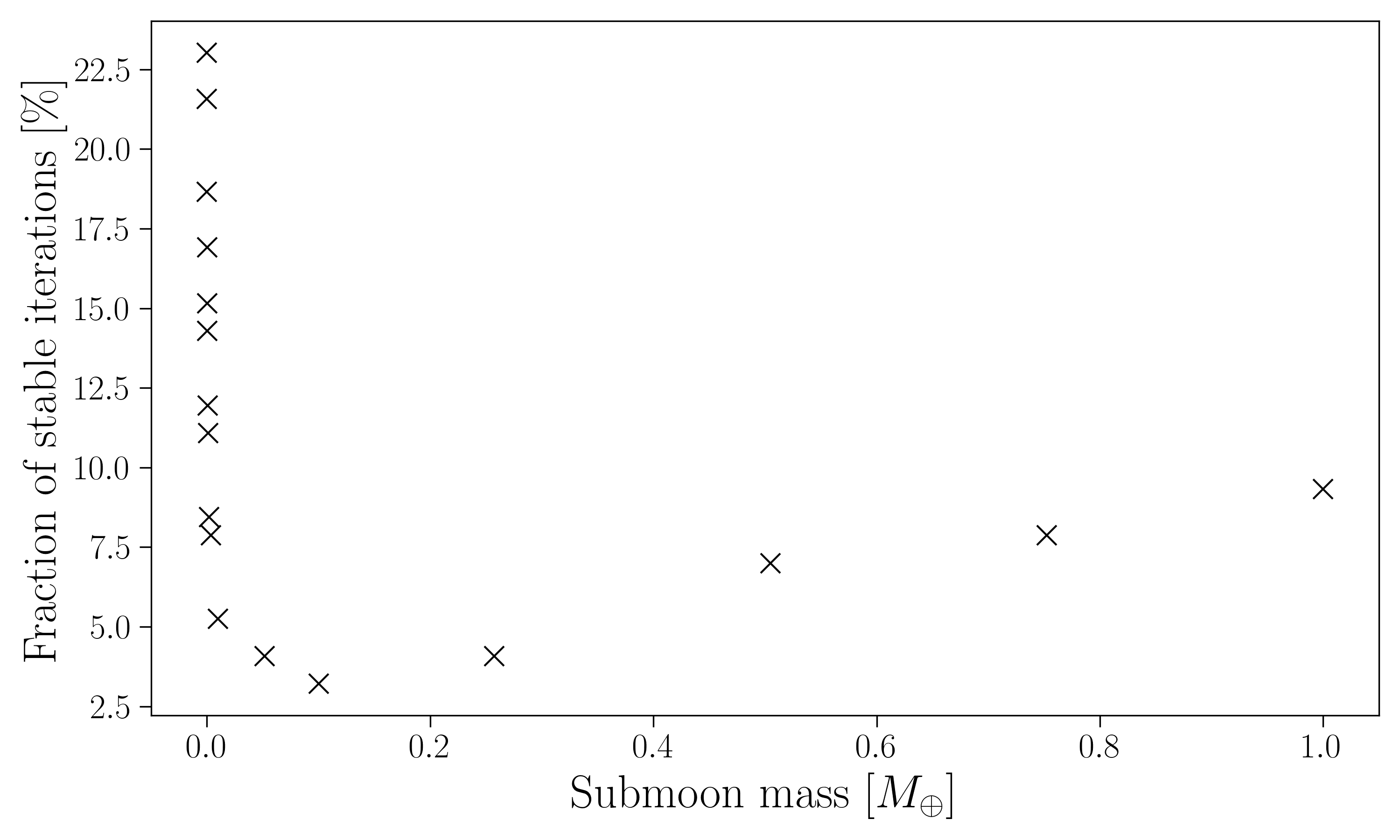}
    \caption{The total fraction of stable iterations in a Warm-Jupiter-like system as a function of submoon mass.
    %The initial semi-major axis of the planet was explored from its Roche-Limit up until $3\mathrm{AU}$, the semi-major axis of the moon and submoon from their respective Roche-Limits to critical semi-major axes. The resolution of the explored semi-major axes parameter space was $(7\times 7\times 7)$.
    }
    \label{fig: Submoon mass unstable valley}
\end{figure}

In this regard, higher submoon masses stabilize the system, since they lead to Roche-zones closer to their primaries and therefore delayed disintegration (inverse relation between satellite mass and Roche-Limit, see Eq. \ref{eq: Roche limit}). On the other hand, much smaller submoon masses lead to smaller tidal forces between the moon and submoon and therefore contribute to the longevity of the system. It seems therefore likely that a "valley" of intermediate submoon masses exists, in which systems are less long-lived. If formation pathways prefer such intermediate mass scales, habitable submoons could be rare objects. Indeed, Fig. \ref{fig: Submoon mass unstable valley} suggests the existence of such a valley with a minimum at around $0.1M_{\oplus}$. The fraction of stable iterations  increases to approximately $18\%$ at $1.8M_{\oplus}$. We don't explore the evolution of the curve thereafter in order to not violate our mass-assumption of $m_{\mathrm{sm}}\leq10\% m_{\mathrm{m}}$. 

\section{Limitations}\label{sec: Limitations}

The tidal model employed in this study is based on the Constant Geometric Lag approximation, wherein the tidal lag angle $\delta_{\text{lag}}$ is assumed to be constant in magnitude (modulo sign), and parameterized via the quality factor $Q$ as $\delta_{\text{lag}}=2\mathrm{sgn}(\Omega - n)/Q$. While this formulation is standard and widely adopted in the literature (e.g., \cite{Murray_Dermott_2000}, \cite{Barnes_2002}, \cite{Sasaki_2012}, \cite{Kollmeier_2018}), it has been shown by \cite{Efroimsky_2013} to be mathematically inconsistent, particularly in its treatment of frequency-dependent dissipation. 

Specifically, the assumption of a constant lag angle is incompatible with a physically realistic rheology, wherein $\delta_{\text{lag}}$ should be a non-constant function of the instantaneous tidal frequency $(\Omega-n)$. In our implementation, we partially mitigate this issue by substituting the discontinuous sign function with a smooth $\mathrm{tanh}(k(\Omega-n))$ transition, introducing an implicit frequency dependence that regularizes the torque near synchronous rotation and enhances numerical stability. Nonetheless, we acknowledge that this modification does not resolve the fundamental limitations of the CGL model.

Our use of it is motivated by the desire to retain analytical tractability and to facilitate a direct extension of the framework developed by \cite{Kollmeier_2018} through the Euler-Lagrange equation. 

While a more physically consistent model would be preferable, incorporating such a formulation would require a substantial rederivation of the governing equations and a full numerical recalibration, which lies beyond the scope of this work. We therefore present our results as a first-order exploratory analysis, with the caveat that future studies should revisit these findings in the context of more rigorous tidal models.

\section{Conclusions}\label{sec: Conclusions}

In this work, we have extended the analytical framework established by \cite{Kollmeier_2018} by incorporating the full tidal interaction network within a nested four-body system: star, planet, moon, and submoon. Our approach is based on deriving the equations of motion from the Euler-Lagrange equation using the Constant Geometric Lag model, enabling us to model tidal dissipation dynamically across all subsystems.

This formulation allowed us to capture key dynamical features absent from the previous analytical model. In particular, our model includes the impact of the planet’s and moon’s migration on the submoon’s lifetime - a mechanism which may either shorten or prolong submoon survival depending on the system’s initial configuration. 

We solved the resulting coupled differential equations numerically by posing an initial value problem across a grid of plausible semi-major axes and spin frequencies. To overcome the numerical instability associated with the discontinuous sign function in the tidal torque, we proposed a smooth approximation using a hyperbolic tangent function. We verified that this approximation agrees well with the original formulation in non-stiff cases and allows us to robustly continue stiff simulations to the full evolutionary timescale of $4.5\mathrm{Gyr}$.

In an Earth-like system (solar-mass star, Earth-mass planet, and Luna-mass moon), we find that a submoon with a mass of up to $4.6\cdot 10^{17}\mathrm{kg}$ (comparable to a large asteroid) could have remained gravitationally stable if the moon migrated to orbit slightly beyond ($a_m\approx 80R_p$) the current Earth-Luna distance ($a_m\approx 60R_p$). This refines the analytical estimate of \cite{Kollmeier_2018} and places an upper bound on submoon survival in terrestrial systems that are near to Earth’s present-day orbital architecture. 

However, for submoons to reach sizes potentially capable of hosting life, much more massive systems are required.

We explored such scenarios in a Kepler-1625-like configuration consisting of a warm Jupiter orbited by a Neptune-mass moon. In this context, submoons as massive as $1.8M_{\oplus}$ - super-Earth class bodies - can survive over billions of years, provided the moon’s orbit is sufficiently wide ($\geq 100R_p$). This implies that massive submoons in exomoon systems around gas giants may offer long-lived environments suitable for habitability.

Interestingly, when reducing the submoon mass to $0.5M_{\oplus}$ - a threshold proposed for sustaining plate tectonics and a long-lived atmosphere \citep{Raymond_2007}  - the overall number of stable configurations decreased. This reflects a trade-off: higher mass submoons benefit from a Roche limit that lies closer to the host, reducing the risk of tidal disintegration, whereas lower mass submoons experience weaker tidal torques and slower migration. Between these regimes, we observe a “valley” of minimal stability at $0.1M_{\oplus}$, where systems are most prone to tidal loss. If submoons in Warm-Jupiter-like systems tend to form near this mass scale, a natural consequence of tidal evolution could be their rarity.

These results place quantitative limits on the lifetimes and orbital niches of submoons, with direct implications for their astrobiological potential. In the most favorable systems - wide-orbiting giant exomoons around warm Jupiters - stable submoons with Earth-like masses may persist for billions of years. Given sufficient energy sources, such as tidal heating or stellar irradiation, these environments could sustain liquid water and tectonic recycling, two key ingredients for habitability.

Having derived the relevant equations from first principles, our framework is well-positioned for extension to more general orbital configurations (e.g., eccentric, inclined, or retrograde orbits) and alternative tidal dissipation models. Future work could incorporate the Constant Time Lag Model to refine torque evolution and assess the impact of eccentricity-driven tidal heating - a factor potentially critical to submoon habitability.

\ack[Acknowledgement]{This research was supported by the
Excellence Cluster ORIGINS, funded by the Deutsche Forschungsgemeinschaft
(DFG, German Research Foundation) under Germany’s Excellence Strategy –
EXC-2094 – 390783311. We acknowledge
the support of the Deutsche Forschungsgemeinschaft (DFG, German
Research Foundation) Research Unit “Transition discs” - 325594231. We thank Leander Ebeling for helpful comments on the manuscript. Portions of the text were refined with the assistance of Large Language Models (ChatGPT-5 by OpenAI) to enhance readability.}

\section{Conflicts of interest}
The authors declare no conflicts of interest.

%%%%%%%%%%%% Biography text %%%%%%%%%%%%%%%%%%%%%%%%%%

\bibliographystyle{apalike}
\bibliography{references}

\end{document}